\documentclass[onecolumn,unpublished,a4paper]{quantumarticle}
\usepackage[utf8]{inputenc}  
\usepackage[backend=biber,style=alphabetic,maxbibnames=999,backref=true]{biblatex}  
\usepackage{amsmath}  
\usepackage{amssymb}  
\usepackage{amsthm}  
\usepackage{amsfonts}  
\usepackage[colorlinks]{hyperref}  
\usepackage[all]{hypcap}  
\usepackage{float}  
\usepackage{listings}  
\usepackage{color}  
\usepackage[T1]{fontenc}  

\DefineBibliographyStrings{english}{
  backrefpage = {Cited on page},
  backrefpages = {Cited on pages},
}

\hypersetup{linkcolor=blue}

\DeclareFixedFont{\ttb}{T1}{txtt}{bx}{n}{4}
\DeclareFixedFont{\ttm}{T1}{txtt}{m}{n}{4}
\definecolor{deepblue}{rgb}{0,0,0.5}
\definecolor{deepred}{rgb}{0.6,0,0}
\definecolor{deepgreen}{rgb}{0,0.5,0}
\newcommand\cppstyle{\lstset{
language=C++,
basicstyle=\ttm,
otherkeywords={uint8_t, __m256i, size_t, ASSERT_TRUE, EXPECT_TRUE, TEST, BENCHMARK},
keywordstyle=\ttb\color{deepblue},
emphstyle=\ttb\color{deepblue},
stringstyle=\color{deepgreen},
commentstyle=\fontfamily{txtt}\selectfont\color{gray},
showstringspaces=false,
literate={*}{{\char42}}1
         {-}{{\char45}}1
}}
\lstnewenvironment{cpp}[1][]
{\cppstyle\lstset{#1}}{}

\newcommand\pythonstyle{\lstset{
language=python,
basicstyle=\ttm,
morekeywords={assert,as,echo},
keywordstyle=\ttb\color{deepblue},
emphstyle=\ttb\color{deepblue},
stringstyle=\color{deepgreen},
commentstyle=\fontfamily{txtt}\selectfont\color{gray},
showstringspaces=false,
literate={*}{{\char42}}1
         {-}{{\char45}}1
}}
\lstnewenvironment{python}[1][]
{\pythonstyle\lstset{#1}}{}

\lstdefinestyle{stimcircuit}{
    language=python,
    basicstyle=\fontsize{6}{6}\selectfont\ttfamily,
    upquote=true,
    stepnumber=1,
    numbersep=8pt,
    showstringspaces=false,
    breaklines=true,
    frame=single,
    aboveskip=1.5em,
    belowskip=1.5em,
    commentstyle=\color{gray},
    classoffset=1,
    morekeywords={DETECTOR,OBSERVABLE_INCLUDE,rec},
    keywordstyle=\color{deepgreen},
    classoffset=2,
    morekeywords={H,R,MPP,M,RX,RY,MY,MX,SQRT\_X,XCY,XCZ,YCX},
    keywordstyle=\color{deepblue},
    classoffset=3,
    morekeywords={X_ERROR,DEPOLARIZE2,DEPOLARIZE1},
    keywordstyle=\color{red},
    classoffset=4,
    morekeywords={TICK,SHIFT_COORDS,QUBIT_COORDS},
    keywordstyle=\color{gray}
}

\renewcommand\thesection{\arabic{section}}

\theoremstyle{definition}

\theoremstyle{definition}

\theoremstyle{definition}

\newcommand{\eq}[1]{\hyperref[eq:#1]{Equation~\ref*{eq:#1}}}
\renewcommand{\sec}[1]{\hyperref[sec:#1]{Section~\ref*{sec:#1}}}
\DeclareRobustCommand{\app}[1]{\hyperref[app:#1]{Appendix~\ref*{app:#1}}}
\newcommand{\fig}[1]{\hyperref[fig:#1]{Figure~\ref*{fig:#1}}}
\newcommand{\tbl}[1]{\hyperref[tbl:#1]{Table~\ref*{tbl:#1}}}
\newcommand{\theoremref}[1]{\hyperref[theorem:#1]{Theorem~\ref*{theorem:#1}}}
\newcommand{\definitionref}[1]{\hyperref[definition:#1]{Definition~\ref*{definition:#1}}}

\begin{document}
\title{New circuits and an open source decoder for the color code}

\date{\today}

\author{Craig Gidney}
\email{craig.gidney@gmail.com}
\affiliation{Google Quantum AI, Santa Barbara, California 93117, USA}

\author{Cody Jones}
\affiliation{Google Quantum AI, Santa Barbara, California 93117, USA}

\begin{abstract}
We present two new color code circuits: one inspired by superdense coding and the other based on a middle-out strategy where the color code state appears halfway between measurements.
We also present ``Chromobius'', an open source implementation of the m{\"o}bius color code decoder.
Using Chromobius, we show our new circuits reduce the performance gap between color codes and surface codes.
Under uniform depolarizing noise with a noise strength of $0.1\%$, the middle-out color code circuit achieves a teraquop footprint of 1250 qubits (vs 650 for surface codes decoded by correlated matching).
Finally, we highlight that Chromobius decodes toric color codes better when given \emph{less} information, suggesting there's substantial room for improvement in color code decoders.
\end{abstract}

\emph{
\textbf{Data availability}:
Chromobius is hosted at \href{https://github.com/quantumlib/chromobius/}{github.com/quantumlib/chromobius} and published \href{https://pypi.org/project/chromobius}{as a pypi package} installed using ``pip install chromobius'' on supported systems.
Circuits generated and stats collected for this paper are available on Zenodo at \href{https://doi.org/10.5281/zenodo.10375289}{10.5281/zenodo.10375289}~\cite{gidney2023colorcodedata}.
}

\tableofcontents
\clearpage

\section{Introduction}

\label{sec:introduction}

\begin{figure}
    \centering
    \resizebox{\linewidth}{!}{
        \includegraphics{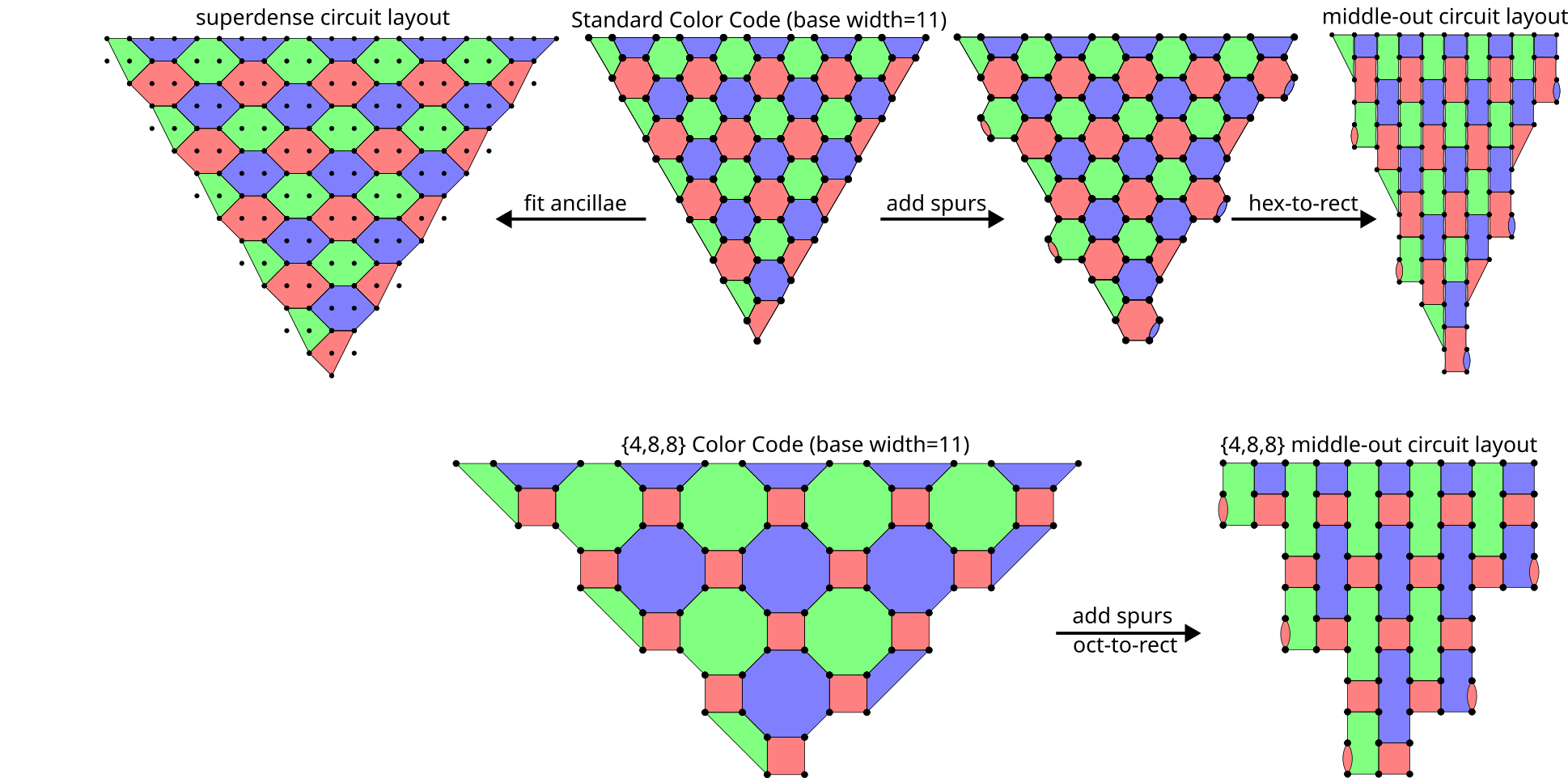}
    }
    \caption{
        Color code layouts.
        Each small black circle is a qubit.
        Each colored shape represents both an X basis stabilizer and a Z basis stabilizer over its vertices.
        Qubits on the vertices of shapes are data qubits; other qubits are measurement ancillae.
        The red, green, and blue colors are a three-coloring of the polygons such that each X error or Z error on a data qubit will flip at most one polygon of each color.
        The logical X (Z) observable is the product of X (Z) on all the data qubits.
        The most common layout is a hexagonal tiling bounded by trapezoids.
        Circuit constructions tweak these layouts to meet various constraints, such as connectivity between ancilla qubits and data qubits.
        For example, the middle-out circuit in this paper adds spurs along the boundaries to avoid stalls during the measurement cycle.
    }
    \label{fig:color-code-layouts}
\end{figure}

The color code is a topological error correcting code, built out of a polygonal tiling where each polygon corresponds to both an X basis and a Z basis stabilizer~\cite{bombin2006colorcode} (see \fig{color-code-layouts}).
The color code is an interesting quantum error correcting code because it shares the lenient quality and connectivity requirements of the surface code~\cite{fowler2012surfacecodereview}, and also has some additional benefits.
For example, like the surface code, the color code can be executed on a planar grid of qubits~\cite{baireuther2019nncolorcode} but, unlike the surface code, the color code has a transversal $S$ gate.

Although having more transversal gates is a nice property, it's likely not the deciding factor when choosing between color codes and surface codes.
During a quantum computation, logical qubits spend a majority of their time idling~\cite{babbush2018,lee2021hypercontraction,gidney2021factor,haner2020improvedeccshot}.
Consequently, a primary factor in deciding between two error correcting codes is how efficiently they \emph{idle}.
This is the comparison we'll focus on in this paper: how many physical qubits are needed to make a good-enough quantum memory.

In terms of efficiently idling, the main difference between the surface code and the color code is that the surface code uses weight 4 stabilizers while the standard color code uses weight 6 stabilizers.
The downside of larger stabilizers is that they are noisier.
Larger stabilizers require more steps to measure and incur more noise per step.
The benefit of larger stabilizers is that they reveal richer information about the errors that occurred.
In particular, basic errors in the the color code produce three detection events instead of two, making errors harder to hide.

In principle, having three detection events per error should allow a more accurate reconstruction of the errors.
Unfortunately, decoding the observed detection events into likely errors has proven to be a difficult problem.
For surface codes, minimum weight matching using the blossom algorithm~\cite{edmonds1965paths} provides an efficient way to find the most likely set of X-type errors that produces a given set of symptoms on the Z-type stabilizers.
No polynomial time equivalent of the Blossom algorithm is known for the color code decoding problem.
This has resulted in a situation where in principle the color code should be capable of outperforming the surface code, but in practice it underperforms the surface code.

A more pragmatic obstacle faced by researchers working on color codes is a lack of tooling.
A researcher exploring the space of surface code circuits has multiple high performance open source decoders to choose between~\cite{higgott2023sparseblossom,wu2023fusionblossom}.
A researcher who wishes to explore the space of color code circuits has to write their own decoder, which can take months.

In this paper, our goal was to explore the space of color code circuits.
To do that, we had to write our own color code decoder, which we've named ``Chromobius".
We're open sourcing Chromobius, in the hopes of reducing the starting penalty for future researchers who want to explore color code circuits.

The paper is structured as follows.
In \sec{circuits}, we describe the two new color code circuits that we wanted to benchmark.
In \sec{chromobius}, we describe the decoder that we implemented to benchmark the circuits.
In \sec{benchmark}, we report the results of benchmarking the circuits and the decoder.
We make closing remarks in \sec{conclusion}.
\app{noise-models} describes the noise models we used for simulations.
\app{euclidean_decoding} describes a variant the color code decoding problem, embedded into the Euclidean plane.
\app{extra-data} includes additional data that didn't fit well in the main paper.

\section{Circuit Constructions}
\label{sec:circuits}

For this paper, we had two ideas for constructing new color code circuits.
The first idea was to use superdense coding to accumulate two measurements at the same time.
The resulting circuit construction is described in \sec{circuits-superdense}.
The second idea was to adapt the no-ancilla middle-out techniques used in \cite{mccewen2023threecouplersurface} from surface codes to color codes.
The resulting circuit construction is described in \sec{circuits-midout}.
We present some extra observations about these circuits in \sec{circuits-distance} and \sec{circuits-pyramid}.

\begin{figure}
    \centering
    \resizebox{\linewidth}{!}{
        \includegraphics{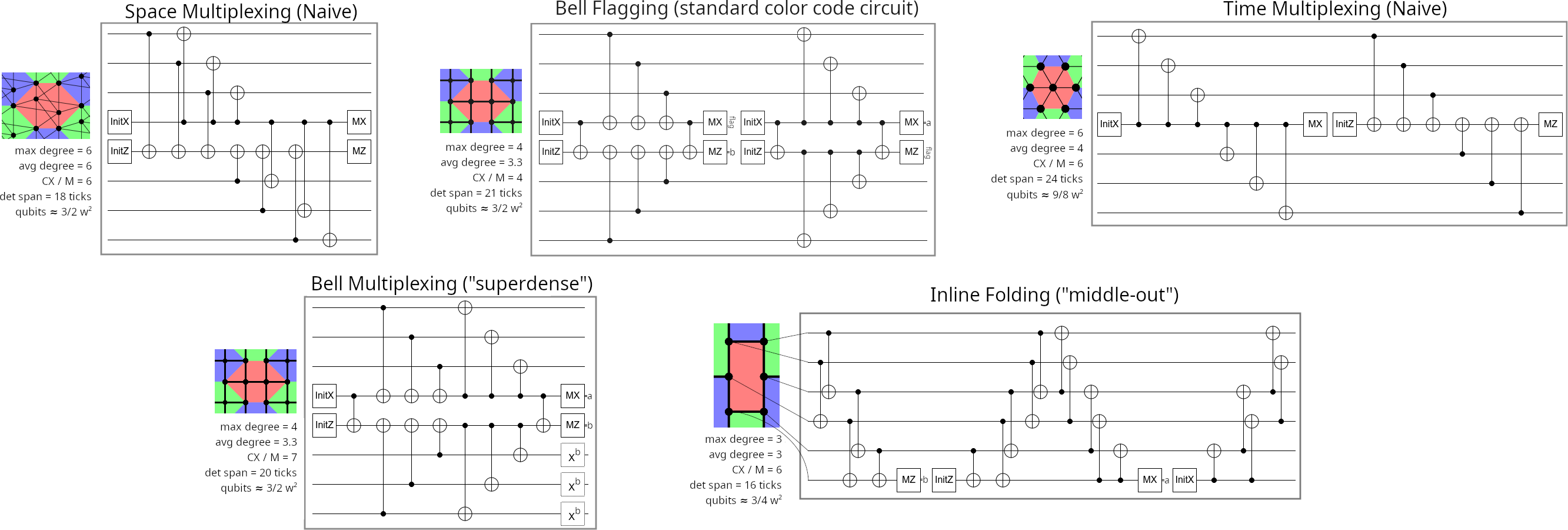}
    }
    \caption{
        Various circuit constructions of the stabilizer measurement cycle for one hexagon within a color code.
        The ``standard color code circuit" is from \cite{baireuther2019nncolorcode}.
        The CX/M number for middle-out circuits is better than the single-hex circuit diagram suggests, because of CX gates being shared by adjacent hexes.
    }
    \label{fig:circuits}
\end{figure}

\subsection{Superdense Color Code Circuits}
\label{sec:circuits-superdense}

Superdense coding~\cite{bennett1992superdense} is a quantum communication protocol that transmits two bits of information by sending one qubit and consuming one previously shared Bell pair.
Superdense coding works because a Bell pair shared between Alice and Bob is stabilized by $+X_A X_B$ and by $+Z_A Z_B$ and the signs of both stabilizers can be negated using either one of the involved qubits.
Alice can negate $+Z_A Z_B$ by applying a bit flip $X_A$, or negate $+X_A X_B$ by applying a phase flip $Z_A$.
Similarly, Bob can negate $+Z_A Z_B$ by applying a bit flip $X_B$, or negate $+X_A X_B$ by applying a phase flip $Z_B$.
This allows two bits to be encoded into the Bell pair, from either side.
Reuniting the Bell pair's qubits allows both bits to be recovered.

Superdense coding's ability to accumulate both bits from both sides is interesting in the context of the color code, where pairs of stabilizers overlap with each other.
Some mechanism is needed to multiplex the measurements of the two overlapping stabilizers onto the available resources.
For example, they could be multiplexed by accumulating the two results onto separate qubits (space multiplexing) or by performing one measurement followed by the other (time multiplexing).
Superdense coding enables multiplexing by \emph{basis}: using controlled bit flips to accumulate one measurement result while using controlled phase flips to accumulate the other.
We call this type of multiplexing ``Bell multiplexing".
See \fig{circuits} for circuit diagrams of various types of multiplexing.

Note that, when using Bell multiplexing, order of operations matters.
It's possible, by reordering the operations accumulating the two measurement results, to kickback CNOTs and other interactions onto the data qubits.
A working ordering is as follows.
Each hexagon contains two measurement ancilla qubits, each connected to three of the hexagon's six data qubits.
The measurement qubits are prepared into a Bell pair.
Then all six data qubits control CNOTs targeting the closest measurement qubit, to accumulate the Z stabilizer measurement.
Then all six data qubits are targeted by CNOTs controlled by the closest measurement qubit, to accumulate the X stabilizer measurement.
Finally, a Bell basis measurement of the measurement qubits is performed.
This measures both of the hexagon's stabilizers, up to Pauli feedback controlled by the measurement results (see \fig{circuits}).
The feedback operations are optimized away at circuit construction time by acting the feedback on the sets of measurements that are compared to produce detection events and observable measurements (e.g. using \href{https://github.com/quantumlib/Stim/blob/main/doc/python_api_reference_vDev.md#stim.Circuit.with_inlined_feedback}{stim.Circuit.with\_inlined\_feedback}).

\begin{figure}
    \centering
    \resizebox{\linewidth}{!}{
        \includegraphics{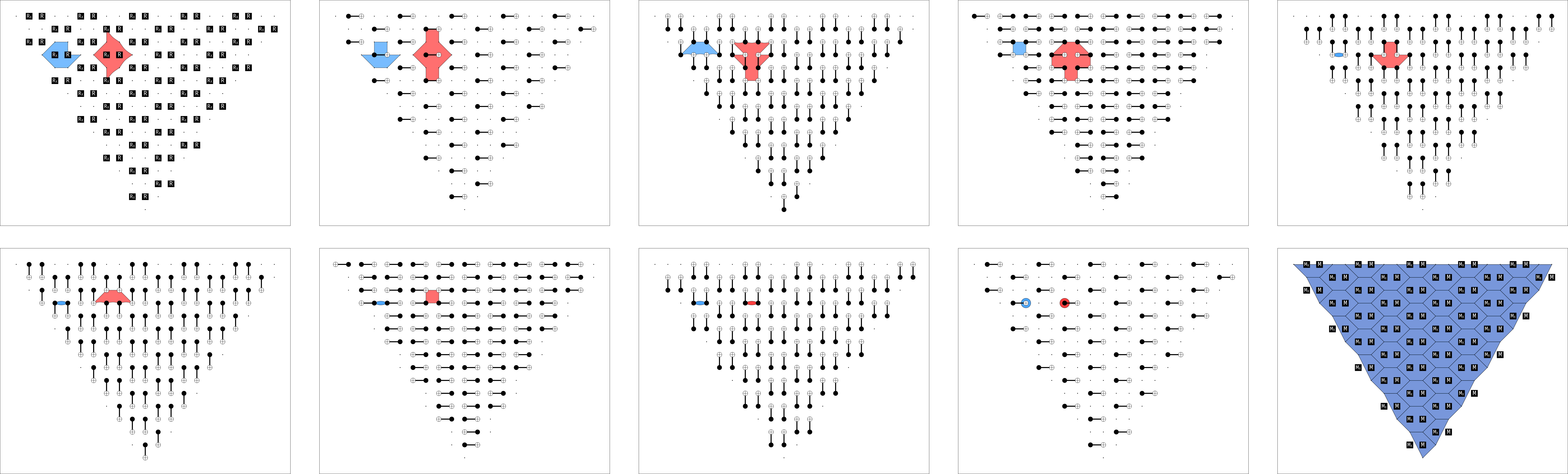}
    }
    \caption{
        One cycle of a superdense color code circuit, including detector slices of a contracting X basis detector (red) and a contracting Z basis detector (blue).
        During the measurement layer, a full detector slice is shown; revealing the color code state.
        The circuit is built by repeating this cycle.
        \href{https://algassert.com/crumble\#circuit=Q(0,0)0;Q(1,0)1;Q(1,1)2;Q(1,2)3;Q(2,0)4;Q(2,1)5;Q(2,2)6;Q(2,3)7;Q(3,0)8;Q(3,1)9;Q(3,2)10;Q(3,3)11;Q(3,4)12;Q(3,5)13;Q(4,0)14;Q(4,1)15;Q(4,2)16;Q(4,3)17;Q(4,4)18;Q(4,5)19;Q(4,6)20;Q(5,0)21;Q(5,1)22;Q(5,2)23;Q(5,3)24;Q(5,4)25;Q(5,5)26;Q(5,6)27;Q(5,7)28;Q(5,8)29;Q(6,0)30;Q(6,1)31;Q(6,2)32;Q(6,3)33;Q(6,4)34;Q(6,5)35;Q(6,6)36;Q(6,7)37;Q(6,8)38;Q(6,9)39;Q(7,0)40;Q(7,1)41;Q(7,2)42;Q(7,3)43;Q(7,4)44;Q(7,5)45;Q(7,6)46;Q(7,7)47;Q(7,8)48;Q(8,0)49;Q(8,1)50;Q(8,2)51;Q(8,3)52;Q(8,4)53;Q(8,5)54;Q(8,6)55;Q(8,7)56;Q(9,0)57;Q(9,1)58;Q(9,2)59;Q(9,3)60;Q(9,4)61;Q(9,5)62;Q(10,0)63;Q(10,1)64;Q(10,2)65;Q(10,3)66;Q(10,4)67;Q(11,0)68;Q(11,1)69;Q(11,2)70;Q(12,0)71;Q(12,1)72;POLYGON(0,0,1,0.5)2_5_10_7;POLYGON(0,0,1,0.5)12_18_26_20;POLYGON(0,0,1,0.5)22_31_42_33_24_16;POLYGON(0,0,1,0.5)28_37_48_39;POLYGON(0,0,1,0.5)44_53_62_55_46_35;POLYGON(0,0,1,0.5)58_64_70_66_60_51;POLYGON(0,1,0,0.5)8_14_22_16_10_5;POLYGON(0,1,0,0.5)24_33_44_35_26_18;POLYGON(0,1,0,0.5)40_49_58_51_42_31;POLYGON(0,1,0,0.5)46_55_48_37;POLYGON(0,1,0,0.5)60_66_62_53;POLYGON(0,1,0,0.5)68_71_70_64;POLYGON(1,0,0,0.5)8_5_2_0;POLYGON(1,0,0,0.5)10_16_24_18_12_7;POLYGON(1,0,0,0.5)40_31_22_14;POLYGON(1,0,0,0.5)26_35_46_37_28_20;POLYGON(1,0,0,0.5)42_51_60_53_44_33;POLYGON(1,0,0,0.5)68_64_58_49;TICK;R_4_6_15_17_19_30_32_34_36_38_50_52_54_56_63_65_67_72;RX_1_3_9_11_13_21_23_25_27_29_41_43_45_47_57_59_61_69_0_2_5_7_8_10_12_14_16_18_20_22_24_26_28_31_33_35_37_39_40_42_44_46_48_49_51_53_55_58_60_62_64_66_68_70_71;MARKX(0)23;MARKZ(1)52;TICK;CX_1_4_3_6_9_15_11_17_13_19_21_30_23_32_25_34_27_36_29_38_41_50_43_52_45_54_47_56_57_63_59_65_61_67_69_72;TICK;CX_2_1_10_9_12_11_22_21_24_23_26_25_28_27_42_41_44_43_46_45_48_47_58_57_60_59_62_61_70_69_5_4_7_6_16_15_18_17_20_19_31_30_33_32_35_34_37_36_39_38_51_50_53_52_55_54_64_63_66_65;TICK;CX_0_1_5_9_7_11_14_21_16_23_18_25_20_27_31_41_33_43_35_45_37_47_49_57_51_59_53_61_64_69_8_4_10_6_22_15_24_17_26_19_40_30_42_32_44_34_46_36_48_38_58_50_60_52_62_54_68_63_70_65;TICK;CX_2_3_8_9_10_11_12_13_22_23_24_25_26_27_28_29_40_41_42_43_44_45_46_47_58_59_60_61_68_69_5_6_14_15_16_17_18_19_31_32_33_34_35_36_37_38_49_50_51_52_53_54_55_56_64_65_66_67_71_72;TICK;CX_1_2_9_10_11_12_21_22_23_24_25_26_27_28_41_42_43_44_45_46_47_48_57_58_59_60_61_62_69_70_4_5_6_7_15_16_17_18_19_20_30_31_32_33_34_35_36_37_38_39_50_51_52_53_54_55_63_64_65_66;TICK;CX_1_0_9_5_11_7_21_14_23_16_25_18_27_20_41_31_43_33_45_35_47_37_57_49_59_51_61_53_69_64_4_8_6_10_15_22_17_24_19_26_30_40_32_42_34_44_36_46_38_48_50_58_52_60_54_62_63_68_65_70;TICK;CX_3_2_9_8_11_10_13_12_23_22_25_24_27_26_29_28_41_40_43_42_45_44_47_46_59_58_61_60_69_68_6_5_15_14_17_16_19_18_32_31_34_33_36_35_38_37_50_49_52_51_54_53_56_55_65_64_67_66_72_71;TICK;CX_1_4_3_6_9_15_11_17_13_19_21_30_23_32_25_34_27_36_29_38_41_50_43_52_45_54_47_56_57_63_59_65_61_67_69_72;TICK;M_4_6_15_17_19_30_32_34_36_38_50_52_54_56_63_65_67_72;MX_1_3_9_11_13_21_23_25_27_29_41_43_45_47_57_59_61_69;MARKX(0)23;MARKZ(1)50_54;TICK;R_4_6_15_17_19_30_32_34_36_38_50_52_54_56_63_65_67_72;RX_1_3_9_11_13_21_23_25_27_29_41_43_45_47_57_59_61_69;MARKX(0)25_21;MARKZ(1)52;TICK;CX_1_4_3_6_9_15_11_17_13_19_21_30_23_32_25_34_27_36_29_38_41_50_43_52_45_54_47_56_57_63_59_65_61_67_69_72;TICK;CX_2_1_10_9_12_11_22_21_24_23_26_25_28_27_42_41_44_43_46_45_48_47_58_57_60_59_62_61_70_69_5_4_7_6_16_15_18_17_20_19_31_30_33_32_35_34_37_36_39_38_51_50_53_52_55_54_64_63_66_65;TICK;CX_0_1_5_9_7_11_14_21_16_23_18_25_20_27_31_41_33_43_35_45_37_47_49_57_51_59_53_61_64_69_8_4_10_6_22_15_24_17_26_19_40_30_42_32_44_34_46_36_48_38_58_50_60_52_62_54_68_63_70_65;TICK;CX_2_3_8_9_10_11_12_13_22_23_24_25_26_27_28_29_40_41_42_43_44_45_46_47_58_59_60_61_68_69_5_6_14_15_16_17_18_19_31_32_33_34_35_36_37_38_49_50_51_52_53_54_55_56_64_65_66_67_71_72;TICK;CX_1_2_9_10_11_12_21_22_23_24_25_26_27_28_41_42_43_44_45_46_47_48_57_58_59_60_61_62_69_70_4_5_6_7_15_16_17_18_19_20_30_31_32_33_34_35_36_37_38_39_50_51_52_53_54_55_63_64_65_66;TICK;CX_1_0_9_5_11_7_21_14_23_16_25_18_27_20_41_31_43_33_45_35_47_37_57_49_59_51_61_53_69_64_4_8_6_10_15_22_17_24_19_26_30_40_32_42_34_44_36_46_38_48_50_58_52_60_54_62_63_68_65_70;TICK;CX_3_2_9_8_11_10_13_12_23_22_25_24_27_26_29_28_41_40_43_42_45_44_47_46_59_58_61_60_69_68_6_5_15_14_17_16_19_18_32_31_34_33_36_35_38_37_50_49_52_51_54_53_56_55_65_64_67_66_72_71;TICK;CX_1_4_3_6_9_15_11_17_13_19_21_30_23_32_25_34_27_36_29_38_41_50_43_52_45_54_47_56_57_63_59_65_61_67_69_72;TICK;M_4_6_15_17_19_30_32_34_36_38_50_52_54_56_63_65_67_72;MX_1_3_9_11_13_21_23_25_27_29_41_43_45_47_57_59_61_69;MARKX(0)23;MARKZ(1)52;TICK;R_4_6_15_17_19_30_32_34_36_38_50_52_54_56_63_65_67_72;RX_1_3_9_11_13_21_23_25_27_29_41_43_45_47_57_59_61_69;TICK;CX_1_4_3_6_9_15_11_17_13_19_21_30_23_32_25_34_27_36_29_38_41_50_43_52_45_54_47_56_57_63_59_65_61_67_69_72;TICK;CX_2_1_10_9_12_11_22_21_24_23_26_25_28_27_42_41_44_43_46_45_48_47_58_57_60_59_62_61_70_69_5_4_7_6_16_15_18_17_20_19_31_30_33_32_35_34_37_36_39_38_51_50_53_52_55_54_64_63_66_65;TICK;CX_0_1_5_9_7_11_14_21_16_23_18_25_20_27_31_41_33_43_35_45_37_47_49_57_51_59_53_61_64_69_8_4_10_6_22_15_24_17_26_19_40_30_42_32_44_34_46_36_48_38_58_50_60_52_62_54_68_63_70_65;TICK;CX_2_3_8_9_10_11_12_13_22_23_24_25_26_27_28_29_40_41_42_43_44_45_46_47_58_59_60_61_68_69_5_6_14_15_16_17_18_19_31_32_33_34_35_36_37_38_49_50_51_52_53_54_55_56_64_65_66_67_71_72;TICK;CX_1_2_9_10_11_12_21_22_23_24_25_26_27_28_41_42_43_44_45_46_47_48_57_58_59_60_61_62_69_70_4_5_6_7_15_16_17_18_19_20_30_31_32_33_34_35_36_37_38_39_50_51_52_53_54_55_63_64_65_66;TICK;CX_1_0_9_5_11_7_21_14_23_16_25_18_27_20_41_31_43_33_45_35_47_37_57_49_59_51_61_53_69_64_4_8_6_10_15_22_17_24_19_26_30_40_32_42_34_44_36_46_38_48_50_58_52_60_54_62_63_68_65_70;TICK;CX_3_2_9_8_11_10_13_12_23_22_25_24_27_26_29_28_41_40_43_42_45_44_47_46_59_58_61_60_69_68_6_5_15_14_17_16_19_18_32_31_34_33_36_35_38_37_50_49_52_51_54_53_56_55_65_64_67_66_72_71;TICK;CX_1_4_3_6_9_15_11_17_13_19_21_30_23_32_25_34_27_36_29_38_41_50_43_52_45_54_47_56_57_63_59_65_61_67_69_72;TICK;M_4_6_15_17_19_30_32_34_36_38_50_52_54_56_63_65_67_72;MX_1_3_9_11_13_21_23_25_27_29_41_43_45_47_57_59_61_69;TICK;R_72_67_65_63_56_54_52_50_38_36_34_32_30_19_17_15_6_4;RX_69_61_59_57_47_45_43_41_29_27_25_23_21_13_11_9_3_1;TICK;CX_69_72_61_67_59_65_57_63_47_56_45_54_43_52_41_50_29_38_27_36_25_34_23_32_21_30_13_19_11_17_9_15_3_6_1_4;TICK;CX_72_71_67_66_65_64_56_55_54_53_52_51_50_49_38_37_36_35_34_33_32_31_19_18_17_16_15_14_6_5_69_68_61_60_59_58_47_46_45_44_43_42_41_40_29_28_27_26_25_24_23_22_13_12_11_10_9_8_3_2;TICK;CX_65_70_63_68_54_62_52_60_50_58_38_48_36_46_34_44_32_42_30_40_19_26_17_24_15_22_6_10_4_8_69_64_61_53_59_51_57_49_47_37_45_35_43_33_41_31_27_20_25_18_23_16_21_14_11_7_9_5_1_0;TICK;CX_65_66_63_64_54_55_52_53_50_51_38_39_36_37_34_35_32_33_30_31_19_20_17_18_15_16_6_7_4_5_69_70_61_62_59_60_57_58_47_48_45_46_43_44_41_42_27_28_25_26_23_24_21_22_11_12_9_10_1_2;TICK;CX_71_72_66_67_64_65_55_56_53_54_51_52_49_50_37_38_35_36_33_34_31_32_18_19_16_17_14_15_5_6_68_69_60_61_58_59_46_47_44_45_42_43_40_41_28_29_26_27_24_25_22_23_12_13_10_11_8_9_2_3;TICK;CX_70_65_68_63_62_54_60_52_58_50_48_38_46_36_44_34_42_32_40_30_26_19_24_17_22_15_10_6_8_4_64_69_53_61_51_59_49_57_37_47_35_45_33_43_31_41_20_27_18_25_16_23_14_21_7_11_5_9_0_1;TICK;CX_66_65_64_63_55_54_53_52_51_50_39_38_37_36_35_34_33_32_31_30_20_19_18_17_16_15_7_6_5_4_70_69_62_61_60_59_58_57_48_47_46_45_44_43_42_41_28_27_26_25_24_23_22_21_12_11_10_9_2_1;TICK;CX_69_72_61_67_59_65_57_63_47_56_45_54_43_52_41_50_29_38_27_36_25_34_23_32_21_30_13_19_11_17_9_15_3_6_1_4;TICK;M_72_67_65_63_56_54_52_50_38_36_34_32_30_19_17_15_6_4;MX_71_70_68_66_64_62_60_58_55_53_51_49_48_46_44_42_40_39_37_35_33_31_28_26_24_22_20_18_16_14_12_10_8_7_5_2_0_69_61_59_57_47_45_43_41_29_27_25_23_21_13_11_9_3_1}{Click here to open a superdense color code circuit in Crumble.}
    }
    \label{fig:superdense-cycle}
\end{figure}

A full cycle of a superdense color code circuit, with feedback inlined, is shown in \fig{superdense-cycle}.
The final circuit is similar to the circuit used in \cite{baireuther2019nncolorcode}, but with flag measurements replaced by the second stabilizer measurements.
Dropping the flags saves four layers of operations per cycle, but reduces the code distance of the circuit.

\subsection{Middle-Out Color Code Circuits}
\label{sec:circuits-midout}

In \cite{mccewen2023threecouplersurface}, various surface code circuits are constructed by starting from a surface code state in the middle of the cycle (halfway between two measurements), instead of the state at the end of the cycle (during the measurements).
The resulting constructions work in an inline fashion, using no ancilla qubits, where stabilizers are folded down to single qubits to be measured and then unfolded back into their original form.
The same ideas can be applied to the color code.

\begin{figure}
    \centering
    \resizebox{\linewidth}{!}{
        \includegraphics{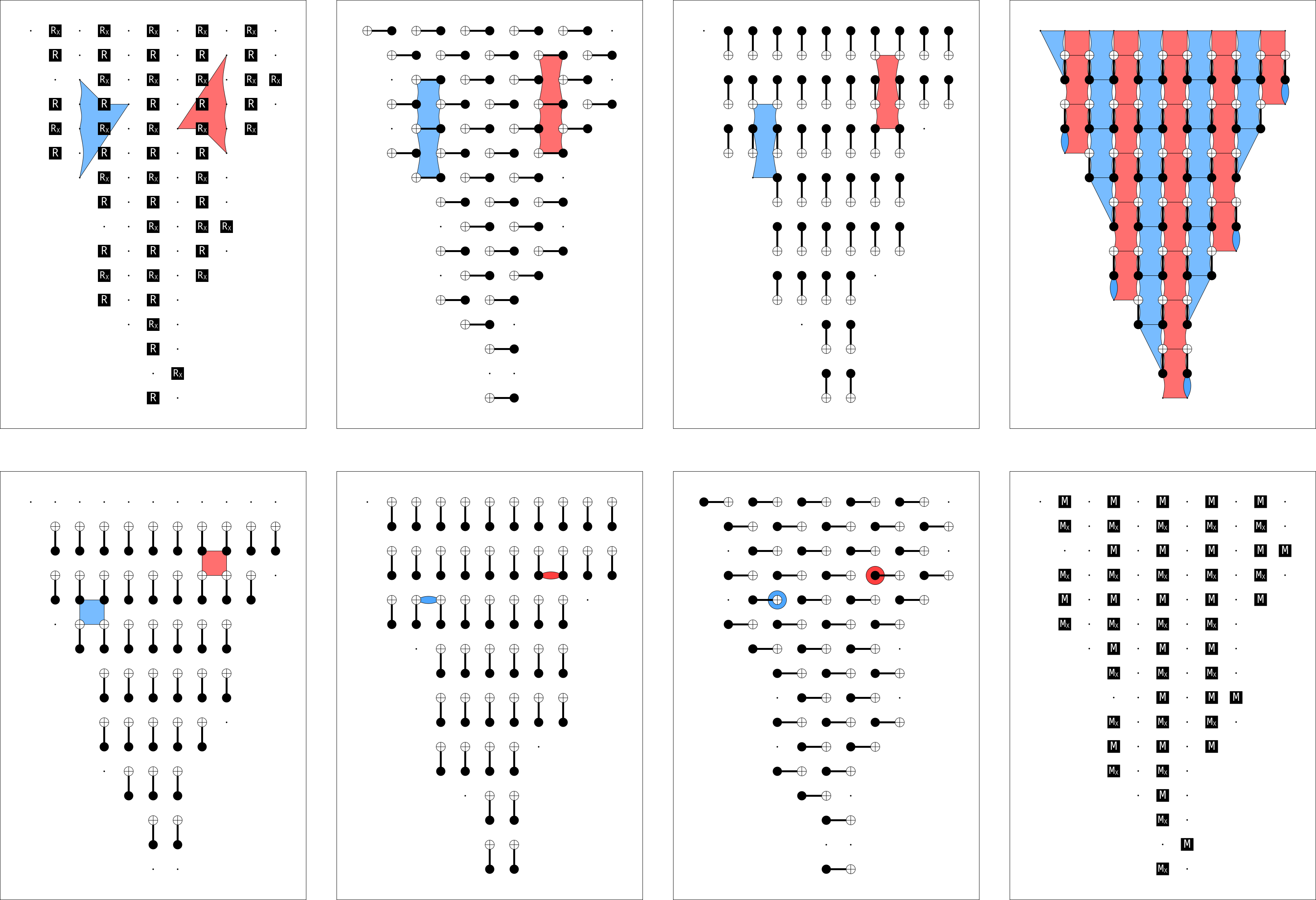}
    }
    \caption{
        One cycle of a middle-out color code circuit, including detector slices of a contracting X basis detector (red) and a contracting Z basis detector (blue).
        At the midpoint a detector slice of all detectors being contracted by this cycle is shown, revealing half of the color code state.
        During the first half of the cycle, the shown stabilizers are transforming from the five body operators of a pyramid code into the six body operators of a color code.
        During the second half of the cycle, the shown stabilizers are contracting into single qubit operators that can be directly measured.
        The full circuit is built by alternating between this cycle and its reverse.
        \href{https://algassert.com/crumble\#circuit=Q(0,1)0;Q(1,1)1;Q(1,2)2;Q(1,3)3;Q(1,4)4;Q(1,5)5;Q(1,6)6;Q(2,1)7;Q(2,2)8;Q(2,3)9;Q(2,4)10;Q(2,5)11;Q(2,6)12;Q(2,7)13;Q(3,1)14;Q(3,2)15;Q(3,3)16;Q(3,4)17;Q(3,5)18;Q(3,6)19;Q(3,7)20;Q(3,8)21;Q(3,9)22;Q(3,10)23;Q(4,1)24;Q(4,2)25;Q(4,3)26;Q(4,4)27;Q(4,5)28;Q(4,6)29;Q(4,7)30;Q(4,8)31;Q(4,9)32;Q(4,10)33;Q(5,1)34;Q(5,2)35;Q(5,3)36;Q(5,4)37;Q(5,5)38;Q(6,1)39;Q(6,2)40;Q(6,3)41;Q(6,4)42;POLYGON(0,0,1,0.5)1_7_8_2;POLYGON(0,0,1,0.5)14_24_25_15;POLYGON(0,0,1,0.5)9_16_17_18_11_10;POLYGON(0,0,1,0.5)19_29_30_31_21_20;POLYGON(0,0,1,0.5)34_39_40_35;POLYGON(0,0,1,0.5)26_36_37_38_28_27;POLYGON(0,0,1,0.75)32_33;POLYGON(0,0,1,0.75)41_42;POLYGON(0,1,0,0.5)0_1_2_3;POLYGON(0,1,0,0.5)4_10_11_12_6_5;POLYGON(0,1,0,0.5)7_14_15_16_9_8;POLYGON(0,1,0,0.5)17_27_28_29_19_18;POLYGON(0,1,0,0.5)13_20_21_22;POLYGON(0,1,0,0.5)24_34_35_36_26_25;POLYGON(1,0,0,0.5)2_8_9_10_4_3;POLYGON(1,0,0,0.5)15_25_26_27_17_16;POLYGON(1,0,0,0.5)11_18_19_20_13_12;POLYGON(1,0,0,0.5)21_31_32_33_23_22;POLYGON(1,0,0,0.5)35_40_41_42_37_36;POLYGON(1,0,0,0.5)28_38_30_29;POLYGON(1,0,0,0.75)5_6;TICK;R_1_5_14_16_18_20_32_34_36_38_41;RX_2_4_6_15_17_19_21_23_35_37_0_3_7_8_9_10_11_12_13_22_24_25_26_27_28_29_30_31_33_39_40_42;MARKX(1)17;MARKZ(0)18;TICK;CX_0_1_2_8_4_10_6_12_7_14_9_16_11_18_13_20_15_25_17_27_19_29_21_31_23_33_24_34_26_36_28_38_35_40_37_42;TICK;CX_2_1_4_3_6_5_8_7_10_9_12_11_15_14_17_16_19_18_21_20_23_22_25_24_27_26_29_28_31_30_33_32_35_34_37_36_40_39_42_41;TICK;CX_3_2_5_4_9_8_11_10_13_12_16_15_18_17_20_19_22_21_26_25_28_27_30_29_32_31_36_35_38_37_41_40;TICK;CX_2_3_4_5_8_9_10_11_12_13_15_16_17_18_19_20_21_22_25_26_27_28_29_30_31_32_35_36_37_38_40_41;TICK;CX_1_2_3_4_5_6_7_8_9_10_11_12_14_15_16_17_18_19_20_21_22_23_24_25_26_27_28_29_30_31_32_33_34_35_36_37_39_40_41_42;TICK;CX_1_0_8_2_10_4_12_6_14_7_16_9_18_11_20_13_25_15_27_17_29_19_31_21_33_23_34_24_36_26_38_28_40_35_42_37;TICK;M_2_4_6_15_17_19_21_23_35_37;MX_1_5_14_16_18_20_32_34_36_38_41;MARKX(1)18;MARKZ(0)17;TICK;R_2_4_6_15_17_19_21_23_35_37;RX_1_5_14_16_18_20_32_34_36_38_41;MARKX(1)18;MARKZ(0)17;TICK;CX_1_0_8_2_10_4_12_6_14_7_16_9_18_11_20_13_25_15_27_17_29_19_31_21_33_23_34_24_36_26_38_28_40_35_42_37;TICK;CX_1_2_3_4_5_6_7_8_9_10_11_12_14_15_16_17_18_19_20_21_22_23_24_25_26_27_28_29_30_31_32_33_34_35_36_37_39_40_41_42;TICK;CX_2_3_4_5_8_9_10_11_12_13_15_16_17_18_19_20_21_22_25_26_27_28_29_30_31_32_35_36_37_38_40_41;TICK;CX_3_2_5_4_9_8_11_10_13_12_16_15_18_17_20_19_22_21_26_25_28_27_30_29_32_31_36_35_38_37_41_40;TICK;CX_2_1_4_3_6_5_8_7_10_9_12_11_15_14_17_16_19_18_21_20_23_22_25_24_27_26_29_28_31_30_33_32_35_34_37_36_40_39_42_41;TICK;CX_0_1_2_8_4_10_6_12_7_14_9_16_11_18_13_20_15_25_17_27_19_29_21_31_23_33_24_34_26_36_28_38_35_40_37_42;TICK;M_1_5_14_16_18_20_32_34_36_38_41;MX_2_4_6_15_17_19_21_23_35_37;MARKX(1)17;MARKZ(0)18;TICK;R_1_5_14_16_18_20_32_34_36_38_41;RX_2_4_6_15_17_19_21_23_35_37;TICK;CX_0_1_2_8_4_10_6_12_7_14_9_16_11_18_13_20_15_25_17_27_19_29_21_31_23_33_24_34_26_36_28_38_35_40_37_42;TICK;CX_2_1_4_3_6_5_8_7_10_9_12_11_15_14_17_16_19_18_21_20_23_22_25_24_27_26_29_28_31_30_33_32_35_34_37_36_40_39_42_41;TICK;CX_3_2_5_4_9_8_11_10_13_12_16_15_18_17_20_19_22_21_26_25_28_27_30_29_32_31_36_35_38_37_41_40;TICK;CX_2_3_4_5_8_9_10_11_12_13_15_16_17_18_19_20_21_22_25_26_27_28_29_30_31_32_35_36_37_38_40_41;TICK;CX_1_2_3_4_5_6_7_8_9_10_11_12_14_15_16_17_18_19_20_21_22_23_24_25_26_27_28_29_30_31_32_33_34_35_36_37_39_40_41_42;TICK;CX_1_0_8_2_10_4_12_6_14_7_16_9_18_11_20_13_25_15_27_17_29_19_31_21_33_23_34_24_36_26_38_28_40_35_42_37;TICK;M_2_4_6_15_17_19_21_23_35_37;MX_1_5_14_16_18_20_32_34_36_38_41;TICK;R_2_4_6_15_17_19_21_23_35_37;RX_1_5_14_16_18_20_32_34_36_38_41;TICK;CX_1_0_8_2_10_4_12_6_14_7_16_9_18_11_20_13_25_15_27_17_29_19_31_21_33_23_34_24_36_26_38_28_40_35_42_37;TICK;CX_1_2_3_4_5_6_7_8_9_10_11_12_14_15_16_17_18_19_20_21_22_23_24_25_26_27_28_29_30_31_32_33_34_35_36_37_39_40_41_42;TICK;CX_2_3_4_5_8_9_10_11_12_13_15_16_17_18_19_20_21_22_25_26_27_28_29_30_31_32_35_36_37_38_40_41;TICK;CX_41_40_38_37_36_35_32_31_30_29_28_27_26_25_22_21_20_19_18_17_16_15_13_12_11_10_9_8_5_4_3_2;TICK;CX_42_41_40_39_37_36_35_34_33_32_31_30_29_28_27_26_25_24_23_22_21_20_19_18_17_16_15_14_12_11_10_9_8_7_6_5_4_3_2_1;TICK;CX_37_42_35_40_28_38_26_36_24_34_23_33_21_31_19_29_17_27_15_25_13_20_11_18_9_16_7_14_6_12_4_10_2_8_0_1;TICK;M_41_38_36_34_32_20_18_16_14_5_1;MX_42_40_39_33_31_30_29_28_27_26_25_24_22_13_12_11_10_9_8_7_3_0_37_35_23_21_19_17_15_6_4_2}{Click here to open a middle-out color code circuit in Crumble.}
    }
    \label{fig:midout-detslice-cycle}
\end{figure}

A two-cycle middle-out circuit for a single hexagon is shown in \fig{circuits}.
To fold a 6-body stabilizer down to a single qubit, we first apply a pair of CNOT gates that fold it into a 4-body stabilizer.
Note that, because each of these CNOT gates runs along two hexagons, in a full color code each CNOT gate is involved in folding two different stabilizers.
Another pair of CNOT gates is used to fold the 4-body stabilizer a 2-body stabilizer.
These CNOTs are also folding two stabilizers instead of one, in a full circuit context.
A final CNOT is used to fold the 2-body stabilizer into a 1-body stabilizer.
(These final CNOTs \emph{aren't} folding two stabilizers at once.)
The 1-body stabilizers can then be measured using normal measurement gates, and the CNOTs can be replayed in reversed order to unfold the stabilizers back into their original form.
This process measures half of the stabilizers of the color code.
The same process then plays out again, but with the roles of the X and Z bases swapped, to measure the other half of the stabilizers.
One cycle of a middle-out circuit over a full color code is shown in \fig{midout-detslice-cycle}.

Because the middle-out circuit has no flag qubits, there are hook errors that expand into multiple data errors.
These hook errors cut the code distance of the circuit in half.
However, the middle-out circuit has benefits that make up for this cost.

The first benefit of the middle-out circuit is that, because it has no ancilla qubits, it can make a bigger color code given the same total number of qubits.
In a color code circuit with two ancilla qubits per hexagon, half the qubits in the system are measurement qubits.
Instead of spending 50\% of its allocated qubits on mitigating hook errors with flags, the middle-out circuit builds a bigger color code.
The middle-out circuit's version of flag qubits is \emph{more color code}.
This recovers a factor $\sqrt{2}$ of the factor $2$ loss in code distance.

The second benefit of the middle-out circuit is its compactness.
Detectors in the middle-out circuit are smaller and thus less noisy than detectors in previous color code circuits.
For example, one way to quantify the size of a detector is to count how many layers of the circuit contain an error that can flip the detector.
Detectors in the middle-out circuit span 16 layers; 20\% less than the other circuits in \fig{circuits}.
Another way to quantify how noisy detectors are is to count the number of CNOT gates per stabilizer measurement.
The middle-out circuit has 6 CNOTs per stabilizer measurement, which is is 15\% less than the other circuits shown in \fig{circuits} (when not counting flag measurements as stabilizer measurements).

The third benefit of the middle-out circuit is its simpler connectivity: a hex grid with 3 neighbors per qubit instead of a square grid with 4 neighbors per qubit.
This has no effect on the simulations we perform, but can be a relevant factor for experimental implementations.
For example, connectivity can affect crosstalk.

\subsection{Distance vs Compactness}
\label{sec:circuits-distance}

Both of the circuit constructions that we have presented fail to achieve the full code distance of the color code.
This is often considered to be a problem, but is something that we actually did intentionally.
Our intuition is that, at physically plausible noise strengths, it's often more important to be compact than to have the best possible code distance.

The source of this intuition is \cite{gidney2022pentagonsurfacecode}, where a circuit was made more compact at the cost of halving the code distance.
The more compact circuit had a better threshold, and performed better at a gate error rate of $0.1\%$, but performed worse for gate error rates of $0.01\%$.
Code distance dominates in the limit of low noise but, for realistic noise, other factors can be more important.
For example, a less compact circuit will have a higher background detection fraction, which reduces the number of additional errors above expectation that are needed to create a logical error.

We'll show in the benchmarking section that, like in \cite{gidney2022pentagonsurfacecode}, our circuits outperform previous work at a noise strength of $0.1\%$ but are worse at a noise strength of $0.01\%$.

\subsection{Pyramid Codes}
\label{sec:circuits-pyramid}

When two codes appear as intermediate states of the same fault tolerant circuit, we refer to these codes as ``cousins''.
Looking at cousins is a useful way to get new perspectives on old codes.
Although cousins may look different, the fact that they can be implemented by the same circuit implies strong similarities.
For example, because the color code has a transversal S gate, its cousins all have a constant-depth fault-tolerant S gate.

\begin{figure}
    \centering
    \resizebox{\linewidth}{!}{
        \includegraphics{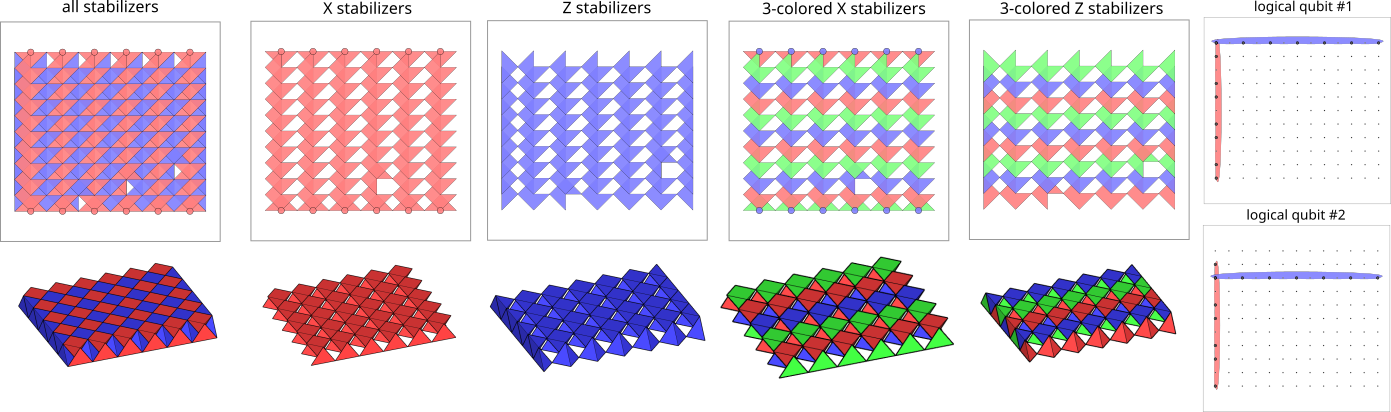}
    }
    \caption{
        The stabilizers of a planar pyramid code.
        Each stabilizer in the bulk is a five-body operator.
        Although the code is defined on a 2d plane, the stabilizers are easiest to understand when drawn as interlocking square-based pyramids with qubits at the vertices of the pyramids.
        Note that this pyramid code has different boundary conditions from the pyramid code that appears during the middle-out color code circuit.
    }
    \label{fig:pyramid-code}
\end{figure}

In the middle-out color code circuit, the color code state appears halfway through the measurement cycle.
At the end of the measurement cycle, a different code appears; one made up of 5-body stabilizers.
We haven't seen this code described before.
We call this 5-body code the ``pyramid code", because the weight-5 stabilizers can be visualized as interlocking square-based pyramids.
See \fig{pyramid-code} for an example of a pyramid code.
Because the pyramid code and the color code both appear as intermediate states of the middle-out circuit, they are cousins.

An interesting property of the pyramid code is that, in each basis, it alternates between color-code-like columns of qubits (where errors cause three vertically-adjacent detection events) and surface-code-like columns of qubits (where errors cause two horizontally-adjacent detection events).
In other words, in a pyramid code, the property that makes decoding color codes difficult is no longer 2D.
It has been squeezed into 1D stripes.
(Maybe this could lead to new ideas for decoding color codes?)

\section{Chromobius}
\label{sec:chromobius}

Chromobius is an implementation of a m{\"o}bius decoder~\cite{sahay2022mobiusdecoder}.
A m{\"o}bius decoder is a color code decoder that works by splitting the decoding problem into three parts: the not-red part, the not-green part, and the not-blue part.
This splitting technique is similar to the one used in restriction decoders~\cite{kubica2019restrictiondecoder}, except that the m{\"o}bius decoder joins the three parts together at the boundaries of the color code.
The benefit of splitting into parts is that the parts can be decoded by a minimum weight matching decoder, and the edges used in the matching can then be lifted into a full solution to the color code problem.
The benefit of joining the parts at the boundaries is that it more accurately weighs the relative costs of matching detection events to a boundary instead of to other detection events in the bulk.

The task of implementing a m{\"o}bius decoder can be split into two main parts: (1) mapping color code errors into m{\"o}bius errors in order to produce the model used by the underlying matching decoder and (2) lifting the solution produced by the matching decoder into a full solution.
Chromobius uses PyMatching~\cite{higgott2023sparseblossom} to solve the matching problems produced by mapping.
Mapping is covered in \sec{chromobius-error-mapping} and lifting is covered in \sec{chromobius-lifting}.
Requirements for these steps to work are detailed in \sec{chromobius-usage}.

\subsection{Mapping Color Code Errors to M{\"o}bius Errors}
\label{sec:chromobius-error-mapping}

Color code errors are mapped into m{\"o}bius errors by splitting each symptom into two parts, and then somehow grouping the split symptoms into pairs.
Each pair will be an edge in the m{\"o}bius matching graph used to configure the minimum weight matcher.
There are four key types of errors that need to be mapped in this way: bulk errors, boundary errors, corner errors, and shift errors.
We'll explain how to map each of these errors into m{\"o}bius errors by example.

A bulk error has one symptom of each color: a red symptom $r$, a green symptom $g$, and a blue symptom $b$.
We describe the error as an instruction ``\texttt{error(p) r g b}'', where $p$ is the probability of the error and $r\ g\ b$ are the symptoms.
To map this error into a matchable error, the red symptom $r$ is copied into the not-green subgraph and the not-blue subgraph of the matching problem (but not the not-red subgraph, since $r$ is red).
This splits $r$ into two symptoms, which we label ``\texttt{r!G}'' and ``\texttt{r!B}''.
Similarly, $g$ splits into ``\texttt{g!R}'' and ``\texttt{g!B}'' and $b$ splits into ``\texttt{b!G}'' and ``\texttt{b!R}''.
Splitting the symptoms has produced a degree 6 error ``\texttt{error(p) r!B r!G g!R g!B b!R b!G}''.
This error can be decomposed into edges by grouping the symptoms by subgraph.
This produces the composite error ``\texttt{error(p) g!R b!R   $\oplus$   r!G b!G   $\oplus$   r!B g!B}'', where ``$\oplus$'' is used as a group separator.
A decoder such as pymatching can approximate this composite error as three independent edge-like error mechanisms ``\texttt{error(p) g!R b!R}'', ``\texttt{error(p) r!G b!G}'', and ``\texttt{error(p) r!B g!B}'', in order to perform matching.

A boundary error has two symptoms, each with a different color.
Consider a boundary error ``\texttt{error(p) r g}'', occurring at the red-green boundary of the color code.
Splitting the symptoms produces the degree 4 error ``\texttt{error(p) r!B r!G g!R g!B}''.
The ``\texttt{r!B g!B}'' symptoms are paired together because they are from the same subgraph.
This leaves the ``\texttt{r!G g!R}'' symptoms, which are paired together because they are all that's left and we don't want any boundary edges (degree 1 errors) in the m{\"o}bius matching graph.
Thus the color code error ``\texttt{error(p) r g}'' becomes the m{\"o}bius matching error ``\texttt{error(p) r!B g!B $\oplus$ r!G g!R}''.
Note how the ``\texttt{r!G g!R}'' component is linking the not-red and not-green subgraphs at the red-green boundary.

A corner error has one symptom.
Consider the corner error ``\texttt{error(p) r}'', occurring at the red corner of a color code.
It splits into the error ``\texttt{error(p) r!G r!B}''.
This error doesn't require decomposing into groups, because it already corresponds to an edge.
However, when testing, we found that squaring the probability of corner errors improved the logical error rate.
So we map ``\texttt{error(p) r}'' to the m{\"o}bius matching error ``\texttt{error(p*p) r!G r!B}'' instead of to ``\texttt{error(p) r!G r!B}''.
(Unlike \cite{sahay2022mobiusdecoder}, we didn't find that manipulating the probabilities of other types of errors was beneficial.)

A shift error has two symptoms of the same color.
For example, measurement errors usually correspond to shift errors.
Consider the shift error ``\texttt{error(p) r1 r2}''.
It splits into the error ``\texttt{error(p) r1!G r1!B r2!G r2!B}''.
These symptoms are grouped by subgraph, producing the m{\"o}bius matching error ``\texttt{error(p) r1!G r2!G $\oplus$ r1!B r2!B}''.

Errors that are combinations of basic errors can also occur.
For example, a Y error on a data qubit in the bulk will produce six symptoms: three that match an X error on that data qubit and three that match a Z error on that data qubit.
The Y error can be mapped into a m{\"o}bius error by realizing it decomposes into these basic X and Z parts, and producing a m{\"o}bius error equal to the combination of the m{\"o}bius errors of its parts.

The error mapping procedure can be somewhat expensive.
However, it's only performed once, when configuring the decoder.
At runtime, when decoding a shot, the work required for mapping is minimal.
The indexing of m{\"o}bius detection events can be arranged such that a color code detection event index $k$ always splits into m{\"o}bius detection events with indices $2k$ and $2k+1$.
So the mapping work done during a shot is just a straightforward doubling of the detection events.

\subsection{Lifting a Matching to a Full Solution}
\label{sec:chromobius-lifting}

To lift a m{\"o}bius matching into a full solution, even under circuit noise, we use a different conceptual strategy than in previous work.
We call this strategy ``tour dragging''.
Tour dragging moves detection events along Euler tours of the matching, merging detection events when they meet and potentially changing them at boundaries.

The solution to a m{\"o}bius matching problem is a set of edges forming paths between detection events.
Consider the subgraph corresponding to just the edges included in the matching.
We can flatten this subgraph by removing the information about which part of the m{\"o}bius matching problem the edge came from.
For example, the edge ``\texttt{r!B g!B}'' and the edge ``\texttt{r!G g!R}'' both flatten into the edge ``\texttt{r g}''.
We call this flattened subgraph the ``flattened matching".

In a flattened matching, all nodes are guaranteed to have even degree.
A valid matching requires that all unexcited nodes are adjacent to an even number of match edges (have even degree in the matching), and all excited nodes are adjacent to an odd number of match edges (have odd degree in the matching)~\cite{higgott2023sparseblossom}.
Flattening unexcited nodes combines two even degree nodes, producing an even degree node.
Flattening excited nodes combines two odd degree nodes, producing an even degree node.

Because all nodes in the flattened matching have even degree, the flattened matching's connected components have Euler tours.
We will solve each connected component separately, and we will solve each component by travelling around its Euler tour while dragging detection events.

When dragging a detection event around a tour, we'll maintain the invariant that the detection event is nearby.
If we're standing on a detector of the same color as the detection event, the detection event will be on that detector.
If the detector is a different color from the detection event, the detection event will be on an adjacent detector.
When the tour reaches a detection event, we grab that detection event.
Anytime we have two detection events, we combine them.
If they have the same color, this cancels them out.
If they have different colors, a bulk error is used to combine them into a single detection event of the remaining color.
When reaching a boundary, we have the option of ignoring the boundary or using it to gain a detection event which can be merged with what we're dragging.
The overall goal is to find a set of actions that produces a self-consistent tour.

For clarity, let's do an example.
Suppose there is a blue detection event ``\texttt{b}'' near a red-green boundary.
The matcher has returned the matching shown in this diagram:

\begin{center}
    \resizebox{0.2\linewidth}{!}{
        \includegraphics{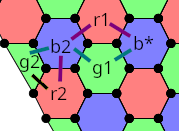}
    }
\end{center}

This matching has seven edges: \texttt{b!G:r1!G}, \texttt{r1!G:b2!G}, \texttt{b2!G:r2!G}, \texttt{r2!G:g2!R}, \texttt{g2!R:b2!R}, \texttt{b2!R:g1!R}, and \texttt{g1!R:b!R}.
They flatten into the Euler tour \texttt{b, r1, b2, r2, g2, b2, g1, [repeat]}.
Following the Euler tour, starting at \texttt{b}, the detection event is grabbed.
Reaching \texttt{r1}, nothing happens because the detection event being dragged is still nearby and is a different color than \texttt{r1} (blue instead of red).
Reaching \texttt{b2}, the detection event needs to be dragged from \texttt{b} to \texttt{b2} because \texttt{b2}'s color matches the dragged detection event's color.
How to do this dragging will have been presolved while configuring the decoder; in this case it's done by inserting ``\texttt{error b2 g1 r1}'' and ``\texttt{error b g1 r1}'' into the set of predicted errors.
Reaching \texttt{g2}, nothing happens because the dragged detection event is still nearby but has a different color.
Crossing to \texttt{r2}, we run into a complication: the g2-to-r2 match edge has contributions from both a bulk error (which preserves net color) and a boundary error (which doesn't).
This means we have a choice: the dragged detection event can either be kept (the bulk case) or discharged (the boundary case).
Chromobius works by tracking both possibilities forwards, ultimately keeping only the one that was self-consistent at the end of the tour.
For this example, the correct choice is to dump the blue excitation into the boundary.
The way to dump the excitation into the boundary will have been presolved at configuration time; in this case it's achieved by inserting ``\texttt{error b2 g2 r2}'' and ``\texttt{error b2 r2}'' into the set of predicted errors.
For the rest of the Euler tour, since no detection event is being dragged, nothing of interest occurs.
At the end of the tour, upon returning to \texttt{b}, we verify that we are not still dragging a detection event.
This certifies that a valid solution was found.

Because of the branching choices available at boundaries, there can in principle be an exponentially large space of solutions for the lifting process to explore.
However, at a given point in the tour there are always only four possible states: dragging no excitation, dragging a red excitation, dragging a green excitation, or dragging a blue excitation.
Therefore a valid solution can be found in linear time by using dynamic programming.
In principle we could attempt to find a \emph{minimal} solution instead of \emph{any} solution, but in most cases all the options will be topologically equivalent.
Also, the task of weighing topologically distinct solutions should already have been done by the matcher, so it would be redundant to do it again.
(Furthermore, in practice, complex cases are exponentially rare because forming large clusters requires many nearby errors~\cite{higgott2023sparseblossom}.)

Beware that there are a variety of corner cases that can occur while dragging detection events around the Euler tours.
For example, a tour may form a ``figure eight'' shape and revisit a node.
When this occurs on a node containing a detection event, a potential bug is for the decoder to pick up the same detection event twice.
Another corner case is that, when crossing a shift error of a different color than the detection event being dragged, the decoder must know how to find a nearby shift error that matches the color of the detection event being dragged in order to keep it near the current location of the tour.

\subsection{Requirements and Assumptions}
\label{sec:chromobius-usage}

Chromobius can be configured using a stim detector error model, which is a representation of a Tanner graph that stim can derive from annotated circuits.
There are four key requirements that a detector error model must meet in order for it to be decodable by Chromobius: \textbf{annotated detectors}, \textbf{rainbow triplets}, \textbf{moveable excitations}, and \textbf{matchable-avoids-color}.

The \textbf{annotated detectors} requirement is that all detectors must be annotated with a basis and color.
This is done via the coordinate data of the detector.
The 4th coordinate of a detector must be set to 0 (basis=X, color=red), 1 (basis=X, color=green), 2 (basis=X, color=blue), 3 (basis=Z, color=red), 4 (basis=Z, color=green), or 5 (basis=Z, color=blue).
For example, the detector error model instruction ``\texttt{detector(0, 0, 0, 5) D10}'' indicates that the detector with index 10 is a blue Z-basis detector.

The \textbf{rainbow triplets} requirement is that, when a basic bulk error has three symptoms, each symptom must have a different color.
If an error with three symptoms repeats a color, Chromobius will attempt to decompose it into other errors.
This can succeed at boundaries, but will fail in the bulk.
Using each color exactly once is important because it ensures bulk errors have neutral charge ($r \equiv g \equiv b \pmod{2}$).

The \textbf{moveable excitations} requirement is that it should be possible to locally solve for how to drag excitations around the bulk.
In practical terms, this means that it should be possible to form shift errors by combining adjacent bulk errors.

The \textbf{matchable-avoids-color} property requires the matchable parts of a circuit to avoid using at least one of the three colors.
For example, if performing lattice surgery between a color code and a surface code, the surface code part should only use one or two colors.
To understand why this is required, suppose that, within the bulk of a matchable region, there were three detectors near each other with each being a different color.
When Chromobius is given a shot with exactly these three detectors activated, the mapping step will produce a m{\"o}bius problem where the not-blue, not-red, and not-green subgraphs each get two of the split detection events.
PyMatching will then, within each subgraph, match the pairs of detection events to each other.
This is a problem because, in a matching problem with an odd number of detection events, at least one must be matched to the boundary.
But the solution found by PyMatching didn't touch any boundaries, meaning the lifting step is guaranteed to fail to find a full solution.

An example of a circuit that meets all these requirements (but isn't a color code) is the circuit for a pyramid code.
A pyramid code may not look like a color code, but its bulk errors use each color exactly once and they pair up into shift errors.
Therefore pyramid codes can be decoded by Chromobius.
An example of a circuit that can't be decoded is the LDPC code from \cite{bravyi2023ldpcbiplanar}.
Although it satisfies the rainbow triplets requirement, it fails to satisfy the moveable excitations property.
Therefore Chromobius cannot decode it.

A surprising example of a circuit that Chromobius can't decode is a color code circuit that cycles between measuring each shape in the X basis, the Y basis, and the Z basis (instead of alternating between X and Z).
These XYZ color code circuits fail to satisfy the moveable excitations property.
The issue is that the XYZ cycle provides no way to locally move an excitation by only 1 round.
Attempting to decode shots from this type of color code with Chromobius can appear to succeed, but soon enough a shot will occur where the matcher finds a path between a blue excitation in round $k$ and an unrelated blue excitation in round $k+1$.
The lifting procedure will then raise an exception, because there's no way to locally merge these excitations.

Another example of a circuit that Chromobius currently can't decode is a color code circuit containing a transversal S gate.
The issue is that the S gate clashes with the need to annotate every detector with a basis.
There needs to be detectors that compare X basis stabilizers before the S gate to the product of X and Z basis stabilizers after the S gate, and these detectors are neither X nor Z basis detectors.
This could be fixed by allowing more complex basis annotations, or improving the code to not require basis annotations in the first place.
Note that it \emph{is} possible to get Chromobius to decode transversal \emph{CNOT} gates in the color code and in the surface code, because CNOTs don't mix the X and Z bases.

Note that Chromobius currently doesn't support flag measurements or heralded errors.
This isn't an inherent limitation of m{\"o}bius decoders, but it wasn't a priority for this paper.
Support for reweighting edges in responds to flags and heralds would be a valuable feature for someone to contribute.
Also note that, for performance reasons, Chromobius doesn't construct a specific set of errors explaining the observed symptoms.
Instead, Chromobius only predicts whether or not annotated logical observables were flipped by those errors.

\section{Benchmarking}
\label{sec:benchmark}

\subsection{Benchmarking Chromobius}

\begin{figure}
    \centering
    \resizebox{0.48\linewidth}{!}{
        \includegraphics{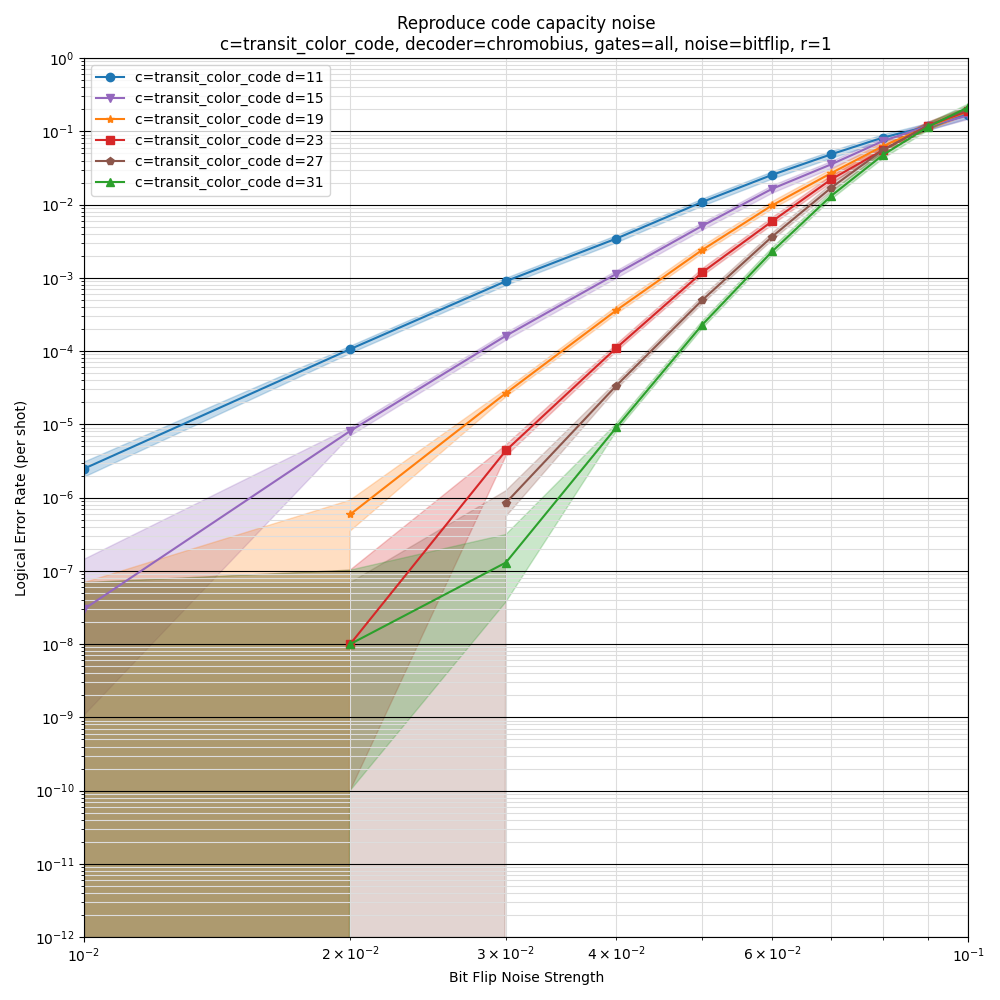}
    }
    \hfill
    \resizebox{0.48\linewidth}{!}{
        \includegraphics{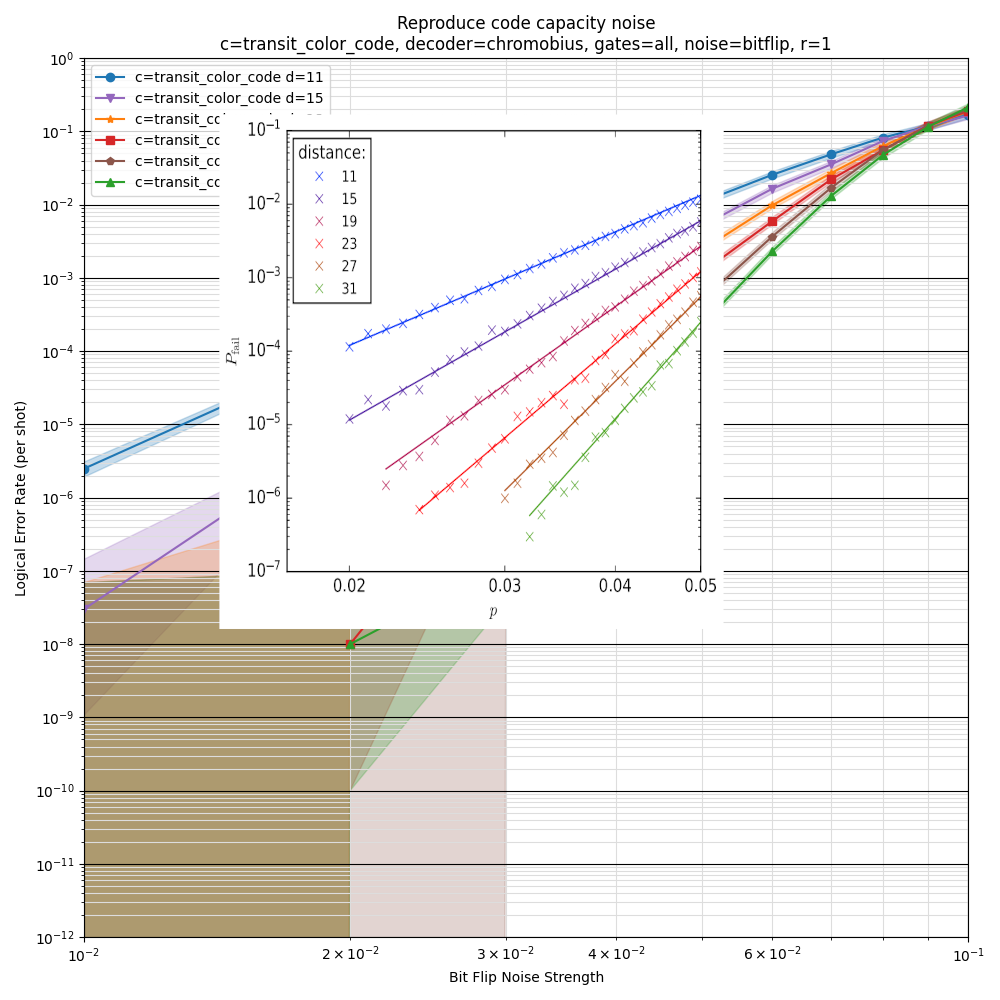}
    }
    \caption{
        Reproducing the code capacity noise results from \cite{sahay2022mobiusdecoder}.
        The left plot shows logical error rate vs bit flip noise strength for a code capacity noise model.
        Highlights show hypotheses with likelihoods within a factor of 1000 of the max likelihood hypothesis, given the sampled data.
        The right plot is the same plot again, but with a copy of figure 11 from \cite{sahay2022mobiusdecoder} overlaid, resized so that the scales line up.
    }
    \label{fig:reproduce}
\end{figure}

To show that Chromobius is behaving as expected for a m{\"o}bius decoder, we reproduced a benchmark from \cite{sahay2022mobiusdecoder}.
In this benchmark, color codes undergo bit flip code capacity noise (see \tbl{noise_model_transit}).
As shown in \fig{reproduce}, our results match the results from \cite{sahay2022mobiusdecoder}.

\begin{figure}
    \centering
    \resizebox{\linewidth}{!}{
        \includegraphics{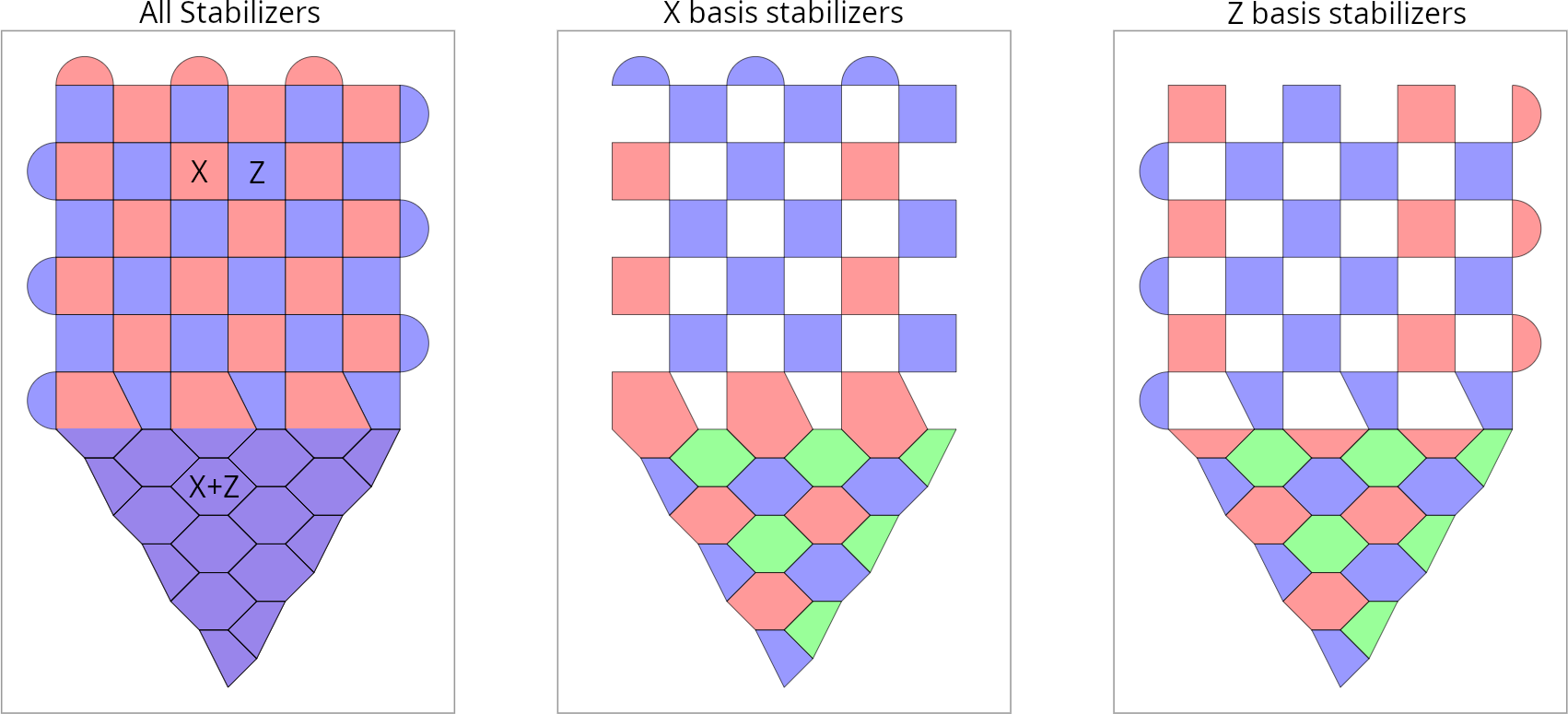}
    }
    \caption{
        Stabilizer configuration of a color code undergoing lattice surgery with a surface code, from \cite{shutty2022colorsurfacemerge}.
        In the left panel, the color of each shape indicates the basis of the corresponding stabilizer(s).
        In the middle and right panels, the colors are the red/green/blue coloring data given to Chromobius.
        Within the surface code part, the coloring satisfies the \textbf{matchable-avoids-color} property by avoiding green.
        The coloring within the surface code part is otherwise arbitrary.
    }
    \label{fig:color2surface}
\end{figure}

\begin{figure}
    \centering
    \resizebox{\linewidth}{!}{
        \includegraphics{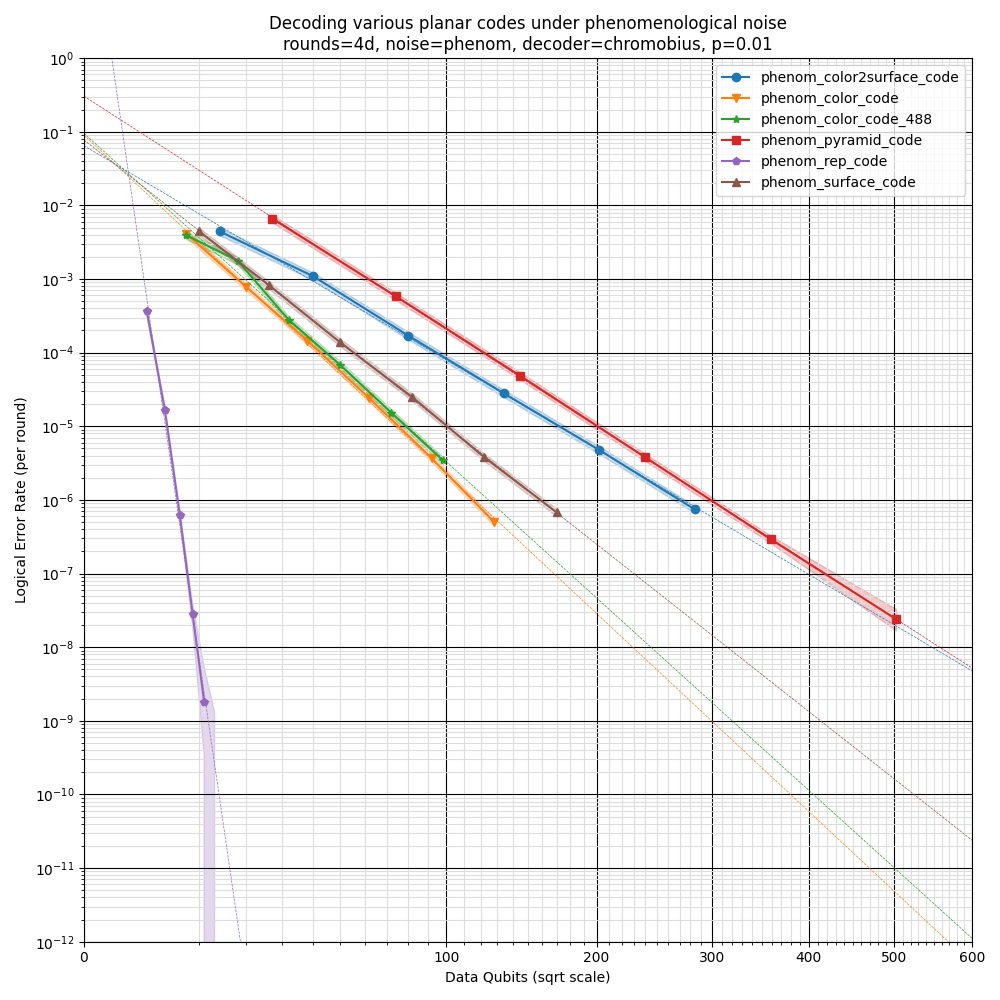}
    }
    \caption{
        Logical error rates of a variety of different codes decoded by Chromobius, under phenomenological noise.
        Highlights show hypotheses with likelihoods within a factor of 1000 of the max likelihood hypothesis, given the sampled data.
    }
    \label{fig:phenom}
\end{figure}

We also benchmarked Chromobius against phenomenological noise (see \tbl{noise_model_phenom}) on a variety of stabilizers codes, to test its ability to generalize beyond a basic color code.
We tested a hexagonal color code, a color code over the $\{4,8,8\}$ tiling, a surface code, a repetition code, a pyramid code, and lattice surgery between a color code and a surface code (from \cite{shutty2022colorsurfacemerge}, see \fig{color2surface}).
The results are shown in \fig{phenom}.

In addition to benchmarking accuracy, we wanted to benchmark speed.
We timed the decoding of superdense color code circuits, middle-out color code circuits, phenomenological color code circuits, code capacity color code circuits, and surface code circuits.
We tried a variety of base widths and noise strengths.
The surface code circuits were decoded using PyMatching, instead of Chromobius, to provide a point of comparison.

Timing was done by invoking decoders from Python.
This has more overhead than using the command line, but is more representative of the typical user experience.
For each circuit, we decoded shots in batches of 1024 until at least one second of decoding work had been done.
If a batch was predicted to exceed the one second target, the batch size was reduced.
All benchmarks were done on an ``n2d-standard-128'' cloud machine.
To reduce noise, only one CPU was used while timing.

\fig{timing_d} highlights some of the collected timing data.
The full data is in the appendix in \fig{timing_d_full} and \fig{timing_r_full}.
Overall, the timing data shows that Chromobius decodes color code detection events at a rate of around 300 kHz, as long as the noise strength is reasonably below threshold.
(This is around an order of magnitude slower than PyMatching decodes surface code detection events.)
The time per detection event increases substantially if the noise strength approaches or passes the threshold.
The time per detection event also increases at extremely low error rates, due to there being so few detection events that fixed costs begin to dominate.

\begin{figure}
    \centering
    \resizebox{\linewidth}{!}{
        \includegraphics{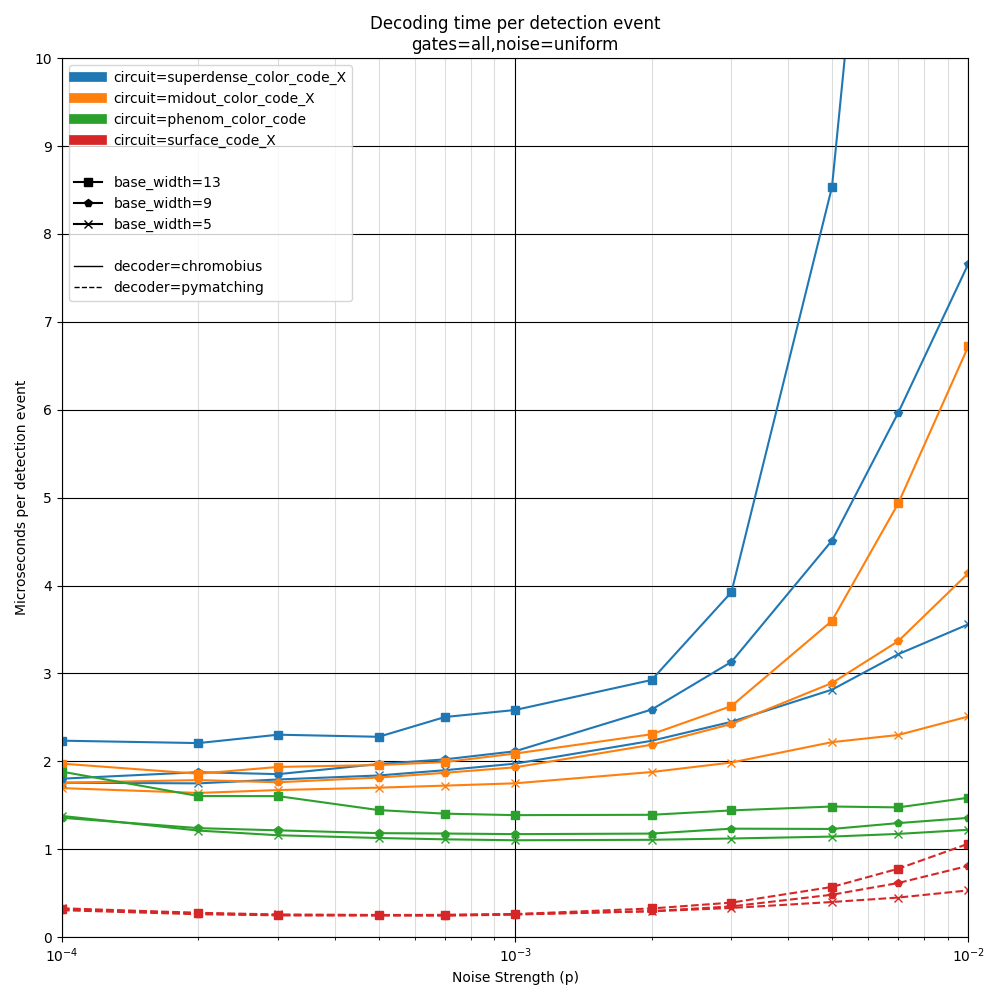}
    }
    \caption{
        Decoding time per detection event for select circuits, sizes, noise strengths, and decoders.
    }
    \label{fig:timing_d}
\end{figure}

\begin{figure}
    \centering
    \resizebox{\linewidth}{!}{
        \includegraphics{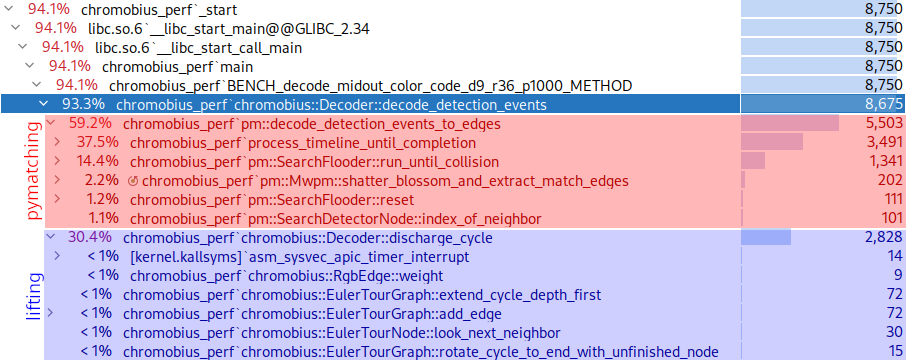}
    }
    \caption{
        Where Chromobius spends its time, according to the Linux ``perf'' tool.
    }
    \label{fig:perf}
\end{figure}

To understand what Chromobius is spending its time on when decoding, we used the Linux ``perf" tool.
The results (see \fig{perf}) indicate that around $60\%$ of decoding time is spent waiting for PyMatching to solve the m{\"o}bius matching problem, and around $30\%$ of decoding time is spent lifting the solution.

\subsection{Benchmarking Circuits}

\begin{figure}
    \centering
    \resizebox{0.48\linewidth}{!}{
        \includegraphics{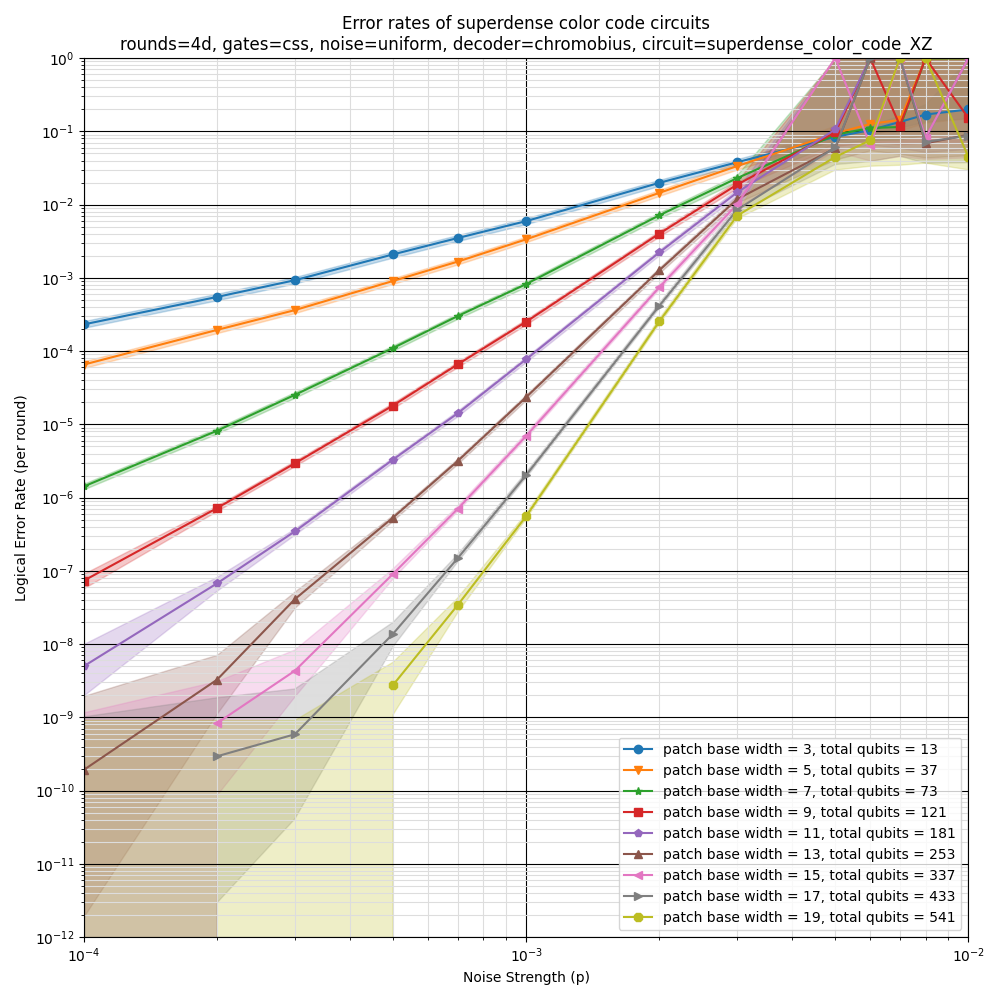}
    }
    \hfill
    \resizebox{0.48\linewidth}{!}{
        \includegraphics{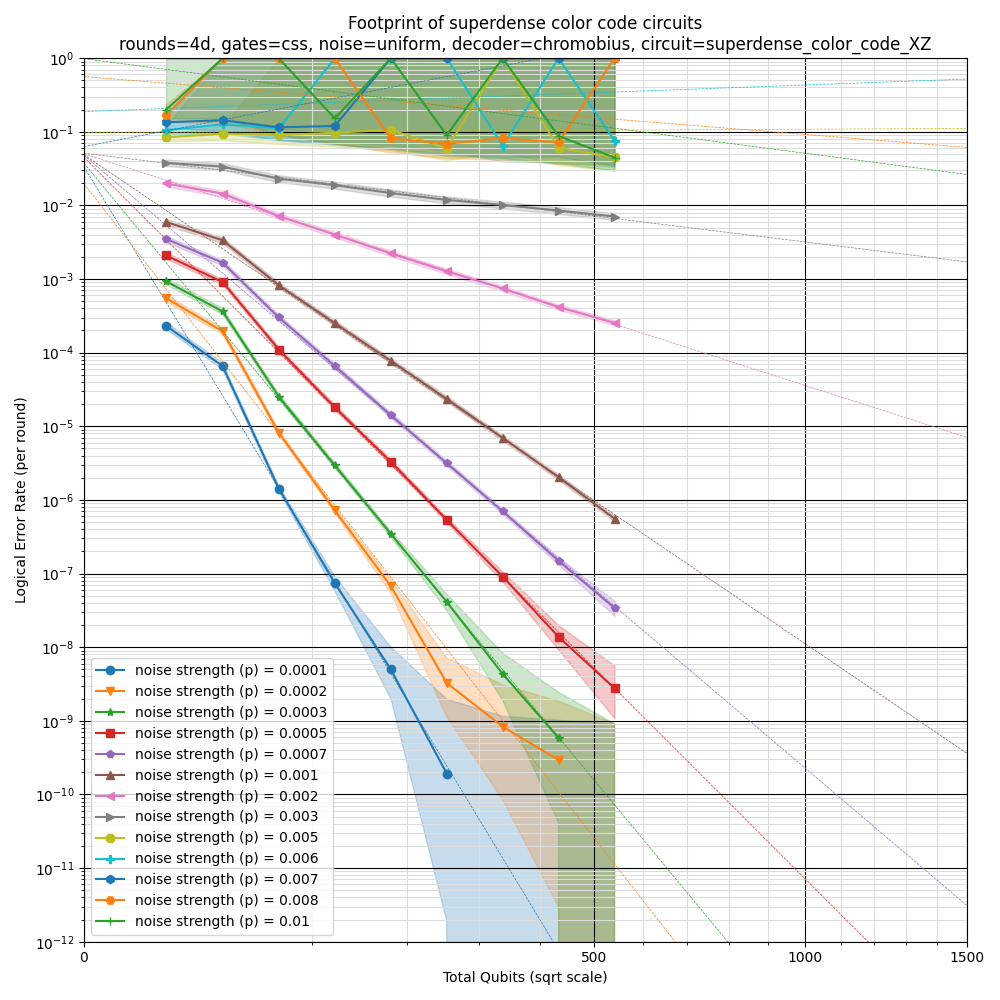}
    }
    \caption{
        Threshold plot (left) and footprint extrapolation plot (right) for superdense color code circuits.
        At a noise strength of $p=0.1\%$, the line fit in the footprint extrapolation plot indicates a teraquop footprint of 2400 physical qubits.
        Highlights show hypotheses with likelihoods within a factor of 1000 of the max likelihood hypothesis, given the sampled data.
    }
    \label{fig:superdense}
\end{figure}

\begin{figure}
    \centering
    \resizebox{0.48\linewidth}{!}{
        \includegraphics{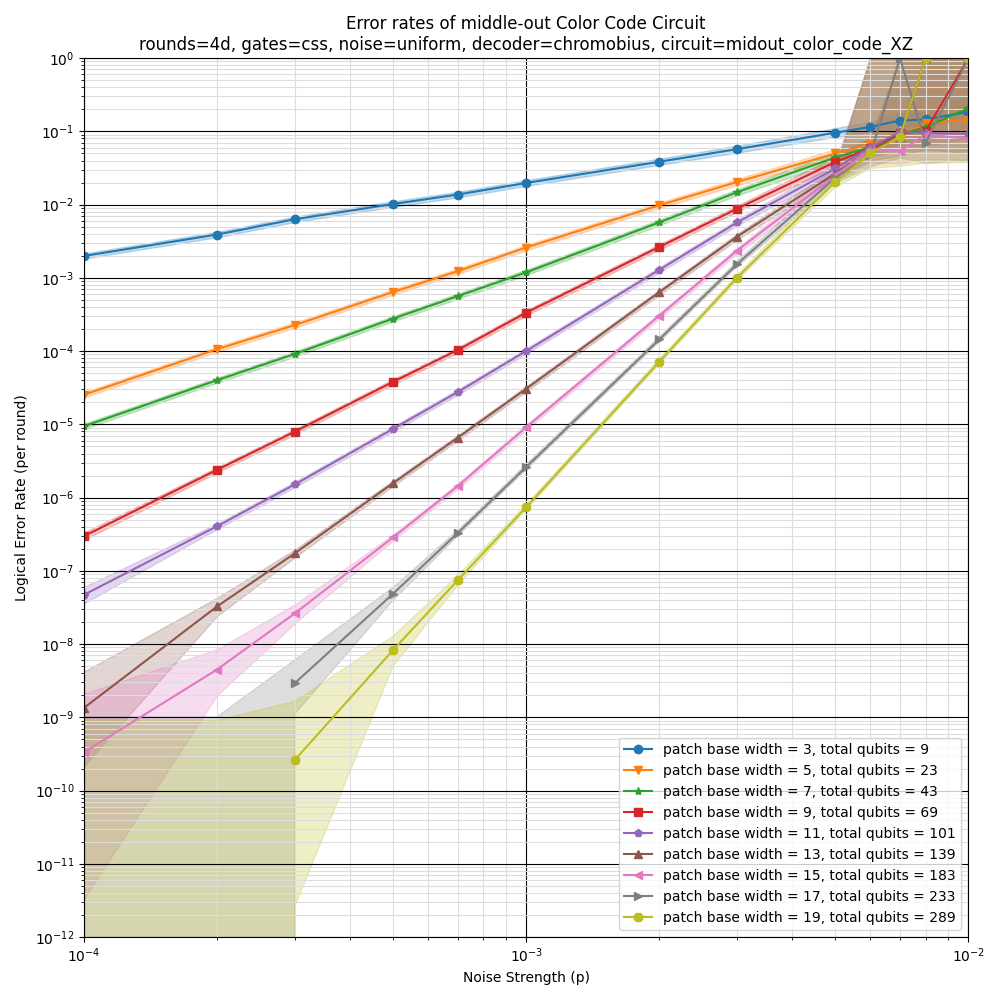}
    }
    \hfill
    \resizebox{0.48\linewidth}{!}{
        \includegraphics{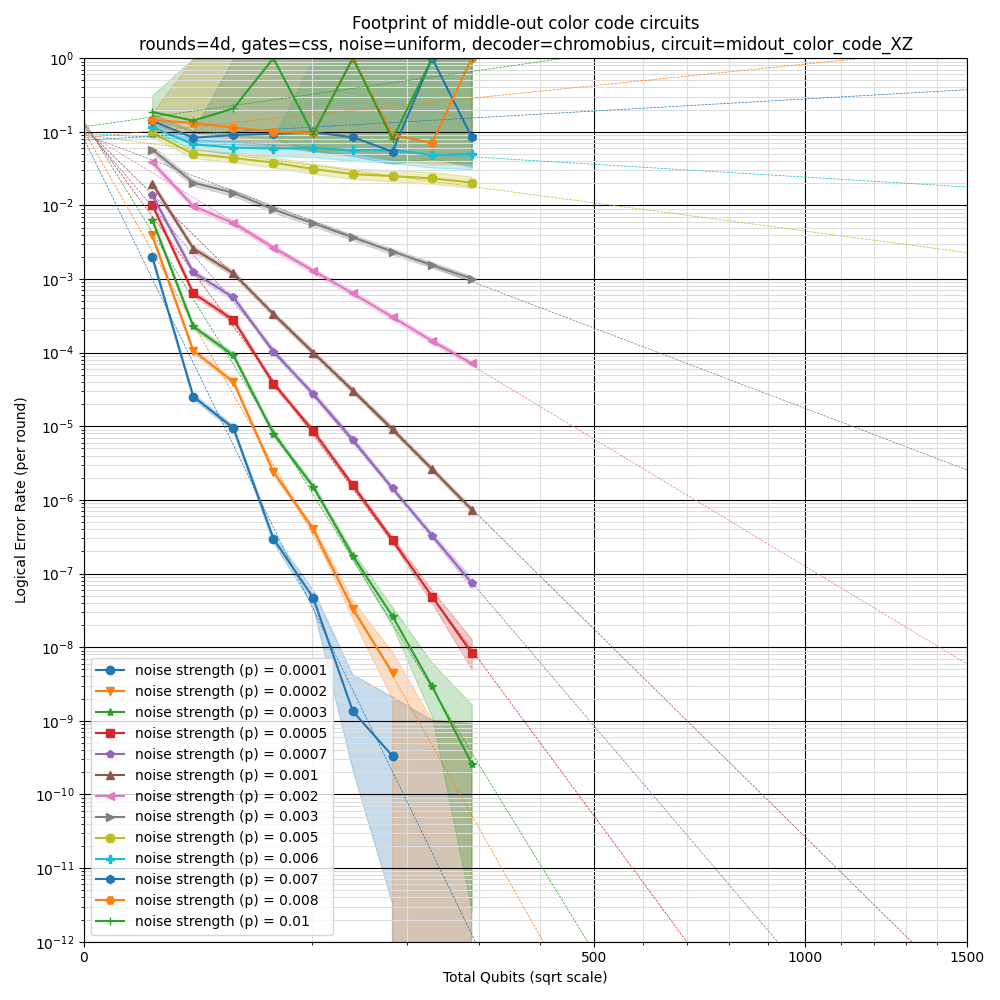}
    }
    \caption{
        Threshold plot (left) and footprint extrapolation plot (right) for the middle-out color code circuit, under uniform depolarizing circuit noise.
        The threshold plot indicates the threshold is between $p=0.5\%$ and $p=0.7\%$.
        At a noise strength of $p=0.1\%$, the line fit in the footprint extrapolation plot indicates a teraquop footprint of 1250 physical qubits.
        Highlights show hypotheses with likelihoods within a factor of 1000 of the max likelihood hypothesis, given the sampled data.
    }
    \label{fig:midout}
\end{figure}

To establish the performance of our circuits, we simulated them preserving both the X and Z observables under uniform depolarizing noise and decoded them using Chromobius.
We tested a variety of sizes and noise strengths.
\fig{superdense} shows the results for superdense color code circuits.
\fig{midout} shows the results for middle-out color code circuits.
These two figures combine the X basis and Z basis simulation results into an overall chance of error by assuming the X and Z observables fail independently.
Plots with data for each individual basis are in the appendix in \fig{superdense-x}, \fig{superdense-z}, \fig{midout-x}, and \fig{midout-z}.

\begin{figure}
    \centering
    \resizebox{\linewidth}{!}{
        \includegraphics{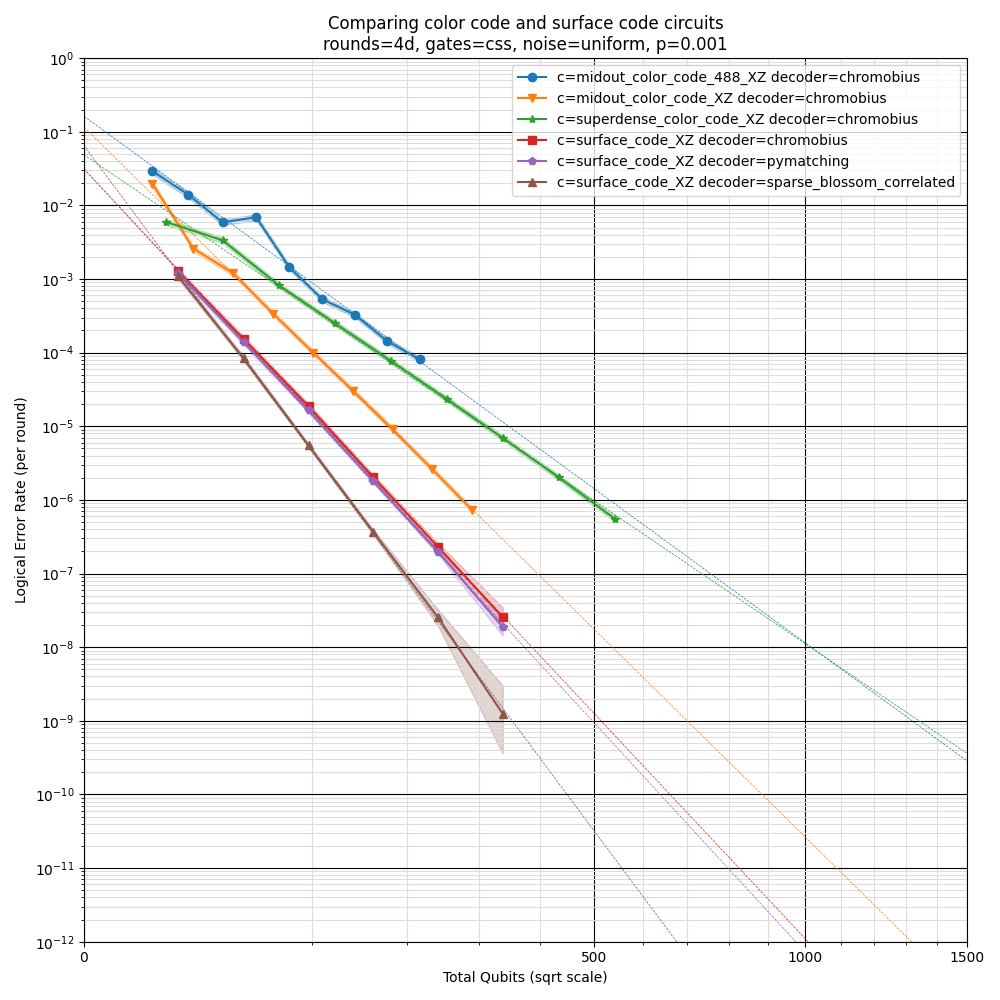}
    }
    \caption{
        Logical error rates of our color code circuit constructions, with the surface code as a reference, versus total number of qubits (including ancilla qubits) used by the circuit.
        The surface code circuits outperform the color code circuits, but the middle-out circuit on a hexagonal lattice is a close competitor.
        Highlights show hypotheses with likelihoods within a factor of 1000 of the max likelihood hypothesis, given the sampled data.
    }
    \label{fig:compare}
\end{figure}

\begin{figure}
    \centering
    \resizebox{0.48\linewidth}{!}{
        \includegraphics{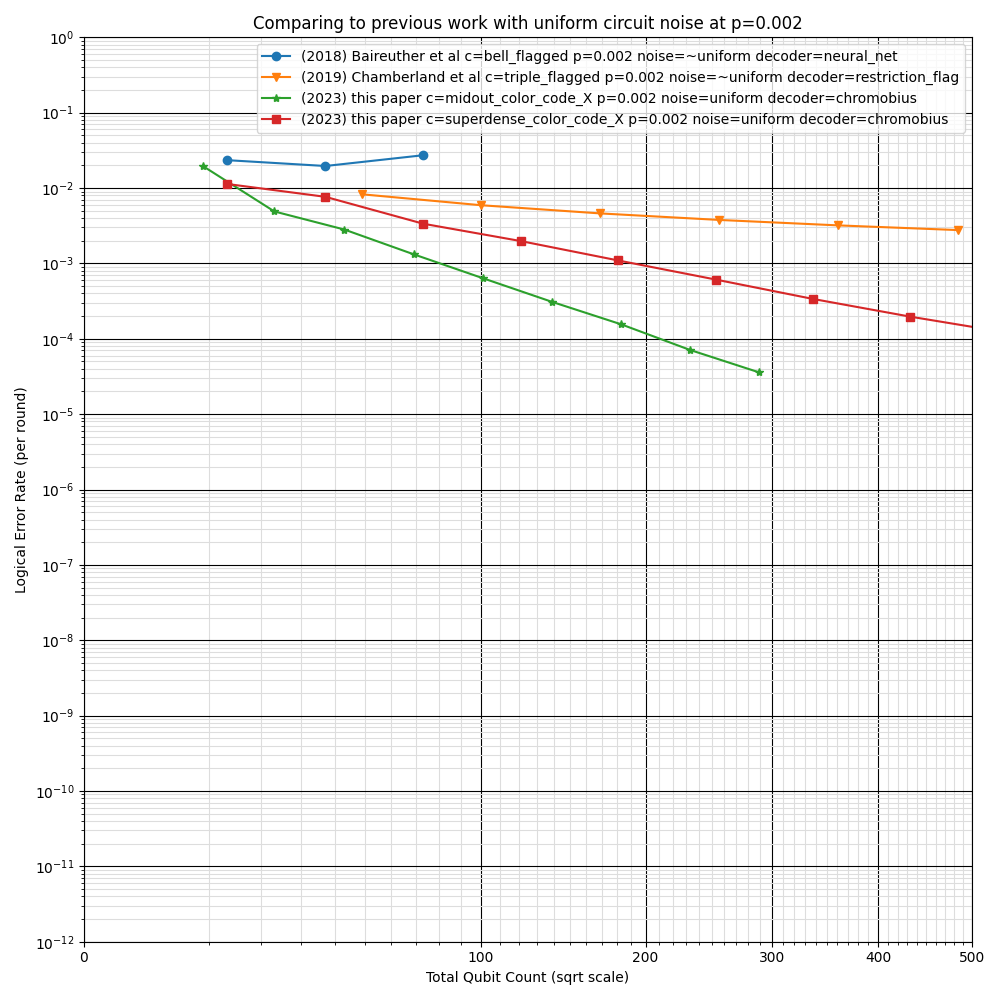}
    }
    \hfill
    \resizebox{0.48\linewidth}{!}{
        \includegraphics{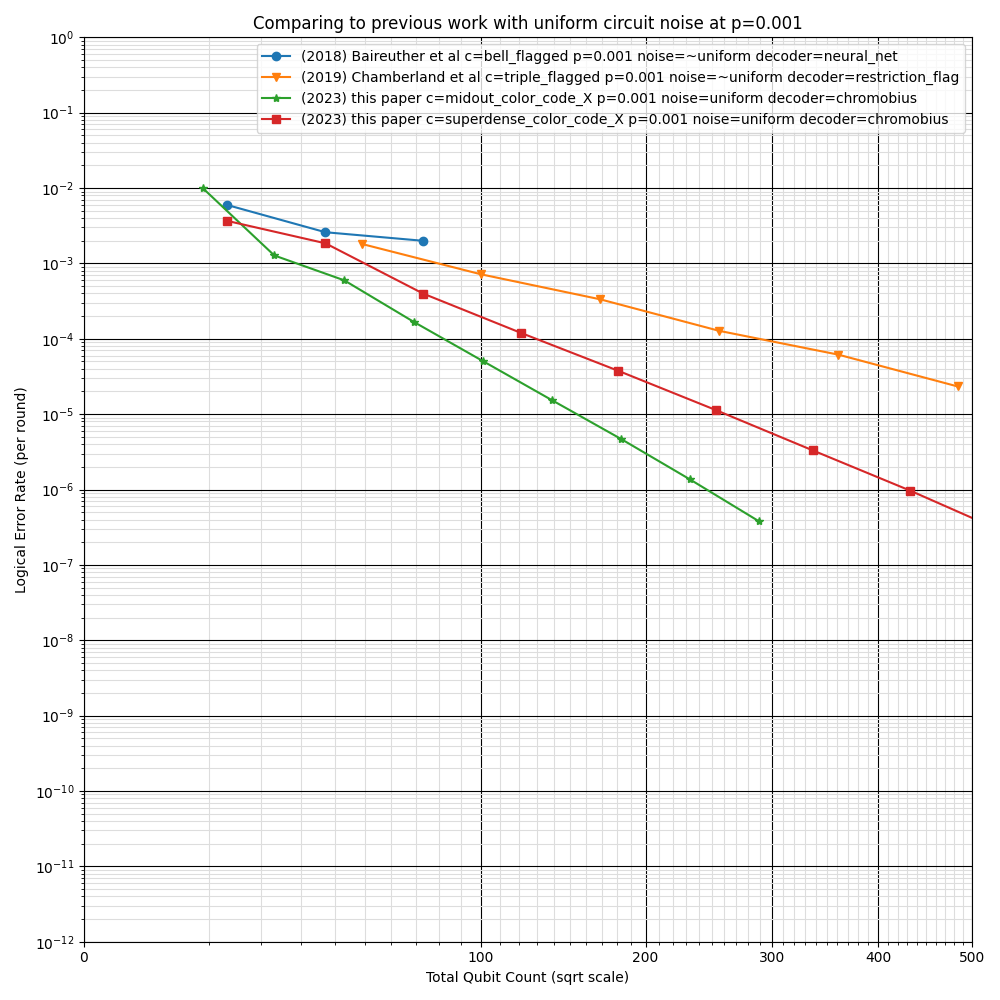}
    }
    \resizebox{0.48\linewidth}{!}{
        \includegraphics{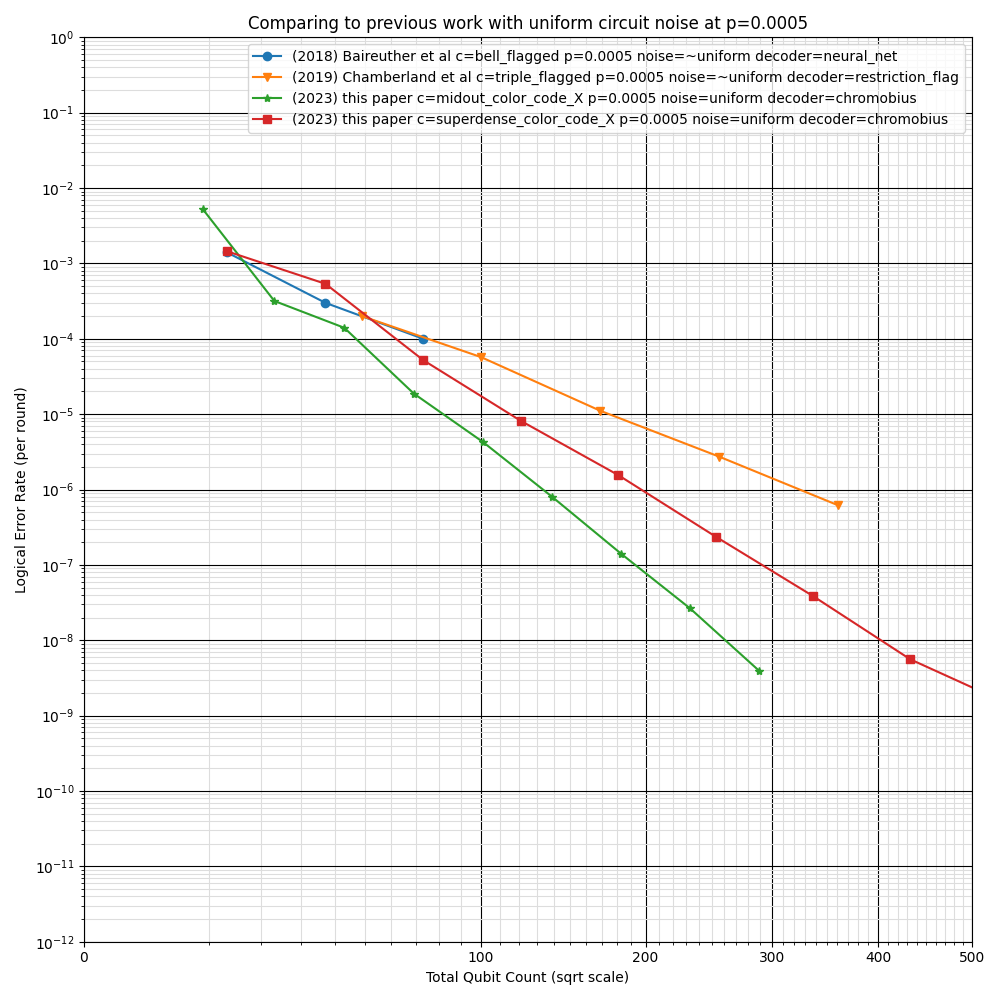}
    }
    \hfill
    \resizebox{0.48\linewidth}{!}{
        \includegraphics{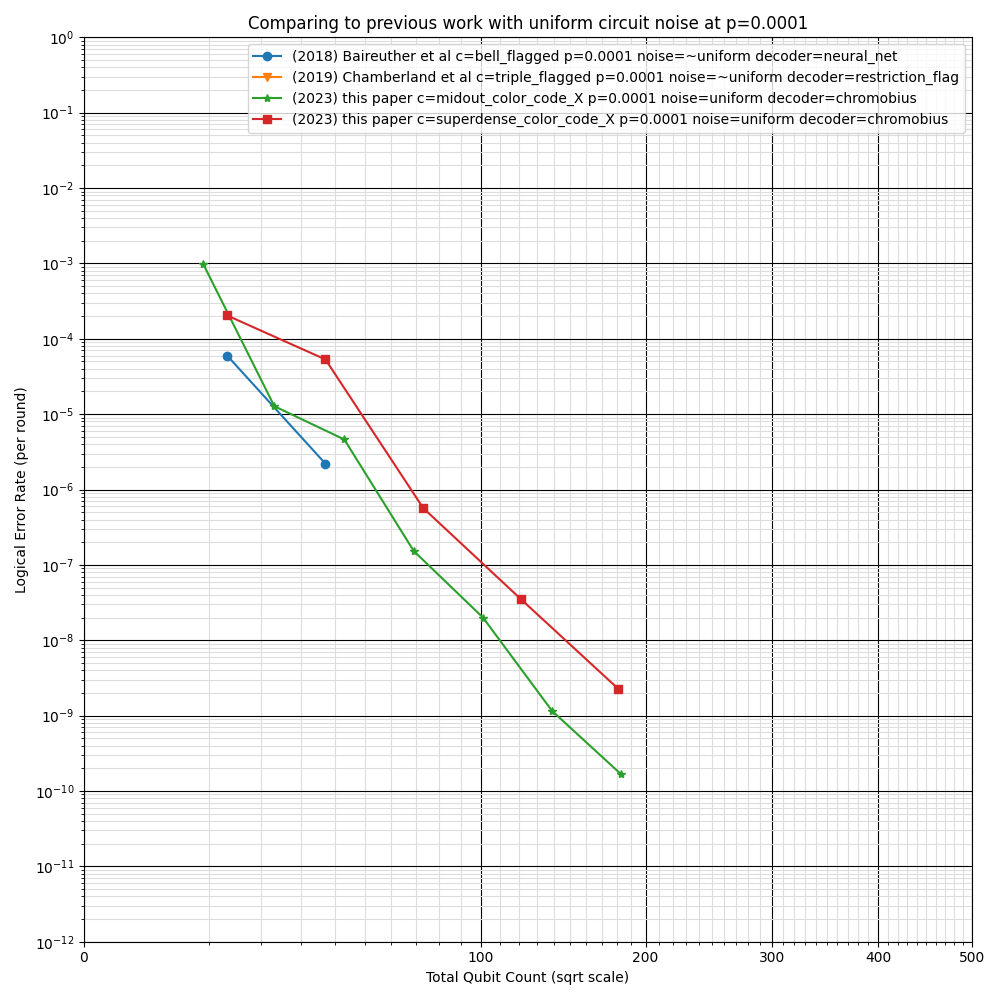}
    }
    \caption{
        Comparing the logical error rates of our circuits to previous work on color code circuits (uncertainties not included).
        Our circuits perform better for noise strengths above $0.02\%$ due to their compactness, and worse for noise strengths below $0.02\%$ due to hook errors halving their code distance.
        The ``Baireuther et al'' results are from figure 4 of \cite{baireuther2019nncolorcode}, adjusted to be per-round instead of per-circuit-layer.
        The ``Chamberland et al'' results are from figure 10 (a) of \cite{chamberland2020triangular}, adjusted to be per-round instead of per-shot.
        The comparison should be reasonably accurate but is not exact due to slight differences in the uniform noise models between papers (e.g. flipping measurement results with probability $2/3\cdot p$ instead of $p$).
        There may also be some slight inaccuracies because we had to eyeball data points, due to a lack of raw data.
    }
    \label{fig:compare_previous_work}
\end{figure}

With this performance data gathered, we were able to compare our circuits to each other, to previous work, and to the surface code.
In \fig{compare_previous_work}, we compare to benchmarks from previous work for other color code circuits.
In \fig{compare}, we compare to the surface code decoded by correlated minimum weight matching.
These figures show that the middle-out circuit outperforms the superdense circuit, and that both circuits outperform previous work at a noise strength of $0.1\%$.
However, this advantage reduces with noise strength; at noise strengths of $0.01\%$ and below our circuits underperform previous work.
Also, our color code circuits still don't outperform the surface code.

The final case that we benchmarked was to investigate an intuition we had about the m{\"o}bius decoder: that it does a good job of understanding the boundaries of the color code, but that it doesn't understand the bulk.
To test this, we simulated toric color codes and ``ablated'' toric color codes.
Toric color codes have no boundaries, so they are ideal for testing a decoder's understanding of the bulk.
Additionally, toric color codes are still fault tolerant (for some choices of observable) when ignoring (ablating) the X-basis stabilizers of one color and the Z-basis stabilizers of a second color.
Toric color codes ablated in this way can be decoded by matching, enabling a reasonably direct comparison between color code decoders and matching-based decoders.
It can also quantify how much useful information a decoder is getting from the ablated stabilizers.

To be clear: in our simulations, the ablated toric color codes perform exactly the same operations as the toric color codes.
They still measure the stabilizers that have been ablated.
They simply ignore the results of the ablated measurements when decoding.
In terms of the stim circuit, this means only DETECTOR annotations are removed.

\begin{figure}
    \centering
    \resizebox{0.7\linewidth}{!}{
        \includegraphics{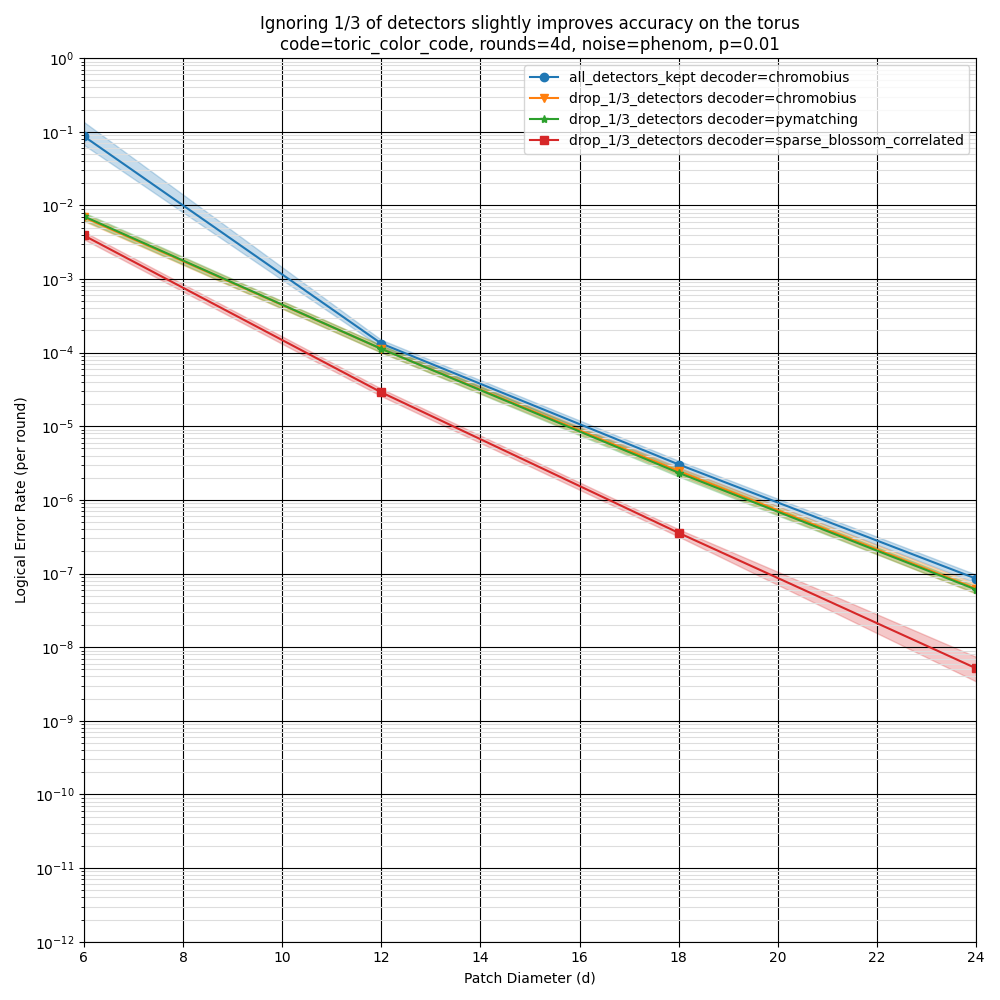}
    }
    \hfill
    \resizebox{0.28\linewidth}{!}{
        \includegraphics{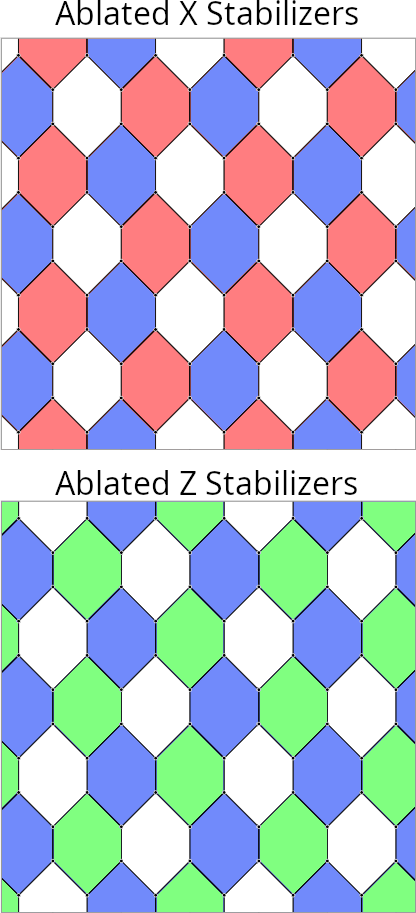}
    }
    \caption{
        Evidence that m{\"o}bius decoders are suboptimal in the bulk of the color code.
        Discarding a third of the stabilizer measurements outperforms not doing that.
        Highlights show hypotheses with likelihoods within a factor of 1000 of the max likelihood hypothesis, given the sampled data.
    }
    \label{fig:hope}
\end{figure}

We decoded toric color codes and ablated toric codes using Chromobius.
We also decoded ablated toric color codes using PyMatching and an internal correlated variant of PyMatching written by Noah Shutty called ``sparse\_blossom\_correlated''.
The results are shown in \fig{hope}.
They show that Chromobius performs slightly better on the \emph{ablated} toric color code.
\textbf{Throwing away a third of the available information made Chromobius give more accurate answers.}
Also note how correlated matching performs even better.
Clearly there's still room for improvement in color code decoders.

\section{Conclusion}
\label{sec:conclusion}

In this paper we presented two new color code circuits, and a new open source color code decoder ``Chromobius".
We showed that Chromobius was fast and accurate: capable of decoding three hundred thousand detection events per second while reproducing the logical error rates from \cite{sahay2022mobiusdecoder}.
We also showed that our circuits achieved lower logical error rates than previous color code circuits, for the same qubit budget, at physically plausible noise strengths.

We hope that Chromobius makes exploring of the space of color code circuits a little bit easier for researchers.
Our results don't achieve the dream of a color code circuit outperforming a surface code circuit, but they do reduce the gap.
Hopefully, future work can finish the job.

\addtocontents{toc}{\protect\setcounter{tocdepth}{-1}}
\section{Contributions}

Craig Gidney wrote Chromobius, designed the middle-out color code circuit, performed the benchmarking, and wrote the paper.
Cody Jones came up with the Bell-multiplexed measurement building block used by the superdense color code circuit.

\section{Acknowledgements}
\addtocontents{toc}{\protect\setcounter{tocdepth}{1}}
We thank Dave Bacon for noticing that the five-body stabilizers of the pyramid code can be visualized as squared-based pyramids.
We thank the Google Quantum AI team for creating an environment where this work was possible.

\printbibliography

\appendix

\clearpage
\section{Noise Models}
\label{app:noise-models}

\begin{table}[H]
    \centering
    \begin{tabular}{|r|l|}
    \hline
    Noise channel & Probability distribution of effects
    \\
    \hline
    $\text{MERR}_B(p)$ & $\begin{aligned}
        1-p &\rightarrow M_{B}
        \\
        p &\rightarrow M_{(-1 \cdot B)} \text{\;\;\;\;\;\emph{(i.e. measurement result is inverted)}}
    \end{aligned}$
    \\
    \hline
    $\text{XERR}(p)$ & $\begin{aligned}
        1-p &\rightarrow I
        \\
        p &\rightarrow X
    \end{aligned}$
    \\
    \hline
    $\text{ZERR}(p)$ & $\begin{aligned}
        1-p &\rightarrow I
        \\
        p &\rightarrow Z
    \end{aligned}$
    \\
    \hline
    $\text{DEP1}(p)$ & $\begin{aligned}
        1-p &\rightarrow I
        \\
        p/3 &\rightarrow X
        \\
        p/3 &\rightarrow Y
        \\
        p/3 &\rightarrow Z
    \end{aligned}$
    \\
    \hline
    $\text{DEP2}(p)$ & $\begin{aligned}
        1-p &\rightarrow I \otimes I
        &\;\;
        p/15 &\rightarrow I \otimes X
        &\;\;
        p/15 &\rightarrow I \otimes Y
        &\;\;
        p/15 &\rightarrow I \otimes Z
        \\
        p/15 &\rightarrow X \otimes I
        &\;\;
        p/15 &\rightarrow X \otimes X
        &\;\;
        p/15 &\rightarrow X \otimes Y
        &\;\;
        p/15 &\rightarrow X \otimes Z
        \\
        p/15 &\rightarrow Y \otimes I
        &\;\;
        p/15 &\rightarrow Y \otimes X
        &\;\;
        p/15 &\rightarrow Y \otimes Y
        &\;\;
        p/15 &\rightarrow Y \otimes Z
        \\
        p/15 &\rightarrow Z \otimes I
        &\;\;
        p/15 &\rightarrow Z \otimes X
        &\;\;
        p/15 &\rightarrow Z \otimes Y
        &\;\;
        p/15 &\rightarrow Z \otimes Z
    \end{aligned}$
    \\
    \hline
    \end{tabular}
    \caption{
        Definitions of noise channels.
        Building blocks for defining noisy versions of gates.
    }
    \label{tbl:noise_channels}
\end{table}

\begin{table}[H]
    \centering
    \begin{tabular}{|r|l|}
    \hline
    Ideal gate & Noisy gate for ``uniform'' model
    \\
    \hline
    $\text{Idle}$ & $\text{DEP1}(p)$
    \\
    $\text{(any single qubit Clifford)} U_1$ & $\text{DEP1}(p) \cdot U_1$
    \\
    $\text{(any two qubit Clifford)} U_2$ & $\text{DEP2}(p) \cdot U_2$
    \\
    \hline
    $R_X$ & $\text{ZERR}(p) \cdot R_X$
    \\
    $R_Z$ & $\text{XERR}(p) \cdot R_Z$
    \\
    \hline
    $M_X$ & $\text{DEP1}(p) \cdot \text{MERR}_X(p)$
    \\
    $M_Z$ & $\text{DEP1}(p) \cdot \text{MERR}_Z(p)$
    \\
    \hline
    \end{tabular}
    \caption{
        The ``uniform'' noise model used by simulations in this paper.
        \tbl{noise_channels} defines each noise channel.
        In legends, circuits experiencing uniform noise are always annotated with ``\_X'' or ``\_Z'' or ``\_XZ''.
        An ``\_XZ'' annotation further indicates the data is the result of combining ``\_X'' and ``\_Z'' statistics under the assumption that the X and Z observables fail independently.
    }
    \label{tbl:noise_model_uniform}
\end{table}

\begin{table}[H]
    \centering
    \begin{tabular}{|r|l|}
    \hline
    Ideal gate & Noisy gate for ``phenom'' model
    \\
    \hline
    $\text{(transition between rounds)}$ & $\text{DEP1}(p)$ \\
    \hline
    $M_{P}$ & $\text{MERR}_P(p)$
    \\
    \hline
    \end{tabular}
    \caption{
        The ``phenom'' noise model used by some simulations in this paper.
        In legends, circuits experiencing uniform noise are always annotated with ``phenom\_''.
        Data qubits are depolarized between rounds.
        Any observable can be measured, but the measurement result is noisy.
        \tbl{noise_channels} defines each noise channel.
        Phenomenological circuits always check that every logical observable is preserved; not just one.
    }
    \label{tbl:noise_model_phenom}
\end{table}

\begin{table}[H]
    \centering
    \begin{tabular}{|r|l|}
    \hline
    Ideal gate & Noisy gate for ``transit bit flip'' model
    \\
    \hline
    $\text{(transition between rounds)}$ & $\text{XERR}(p)$ \\
    \hline
    $M_{P}$ & $M_{P}$
    \\
    \hline
    \end{tabular}
    \caption{
        The ``transit bit flip'' noise model used by some simulations in this paper.
        In legends, circuits experiencing transit bit flip noise are always annotated with ``transit\_''.
        This is a code capacity noise model.
        Between rounds, a probabilistic bit flip is applied to each data qubit.
        Any observable can be measured.
        There are no measurement errors.
        \tbl{noise_channels} defines each noise channel.
        Transit circuits experiencing bit flip noise only check that the logical Z observable is preserved.
    }
    \label{tbl:noise_model_transit}
\end{table}

\clearpage
\section{The Euclidean Color Code Decoding Problem}
\label{app:euclidean_decoding}

In this section we define a continuous version of color code decoding problem, on the Euclidean plane.
This problem should contain the core difficulty that makes color code decoding difficult, but not any of the jagged corner cases of a discretized distance metric.
We include it here because it provides a different perspective on the color code decoding problem.

In a Euclidean Color Code Decoding Problem you are given a set $C$ of colored points where each colored point $(c, p)$ has a color $c \in \{1, 2, 3\}$ and a two dimensional location $p \in \mathbb{R}^2$.
Note that colors always combine by xoring ($\oplus$), and color 0 is neutral.
It's promised that points don't overlap and that the net color is neutral:

$$\bigoplus_{(c, p) \in C} c = 0$$

A solution to the problem is a set of additional neutral points $N$ and a set of colored edges $E$.
Each element of $N$ is a neutral colored point $(c, p)$ where $c=0$, $p \in \mathbb{R}^2$, and $p$ isn't a point used by $C$.
Each element of $E$ is an edge $(c, p_1, p_2)$ where $c \in \{1, 2, 3\}$, $p_1 \in \mathbb{R}^2$, and $p_2 \in \mathbb{R}^2$.

For a solution to be valid, the end points of each edge must terminate on nodes.
Also, the net color of the edges touching a node must equal the node's color.

For an edge $(c, p_1, p_2) \in E$ to be valid, it must terminate on nodes:

$$\left(\exists c_n : (c_n, p_1) \in (C \cup N)\right) \land \left(\exists c_n : (c_n, p_2) \in (C \cup N)\right)$$

For a node $(c, p) \in (C \bigcup N)$ to be valid, it must touch edges explaining its color:

$$c = \bigoplus_{(c_e, p_1, p_2) \in E} \begin{cases}
p = p_1 & c_e\\
p=p_2 & c_e\\
\text{else} & 0\end{cases}$$

If all edges and nodes are valid, then the solution is valid.

The cost of a solution is the sum of the lengths of its edges:

$$\text{cost}(N, E) = \sum_{(c, p_1, p_2) \in E} \text{EuclideanDistance}(p_1, p_2)$$

The goal of the decoding problem is to find a valid solution with the lowest cost.

The Euclidean color code decoding problem is superficially similar to the Steiner tree problem~\cite{wikipedia_steiner_tree}, which is known to be NP-complete.
However, there are two key differences.
First, the Steiner tree problem has no notion of ``color''.
There is no constraint where the edges connected to a virtual node added by the solution must have colors that sum to neutral.
Second, the Steiner tree problem is about creating connections whereas the color code decoding problem is about neutralizing charges.
In particular, the Steiner tree problem requires that the solution has only one connected component.
In the color code decoding problem, the solution can have multiple connected components as long as each component has a net neutral charge.

\begin{figure}[H]
    \centering
    \resizebox{0.4 \linewidth}{!}{
        \includegraphics{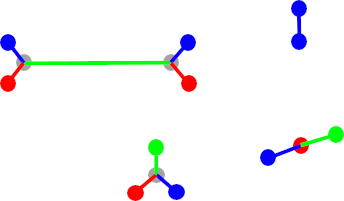}
    }
    \caption{
        Example of a solved Euclidean color code decoding problem.
        Color 1 is red, color 2 is green, and color 3 is blue.
        Red, green, and blue circles are the input colored points $C$ defining the problem.
        Red, green, and blue edges are the colored edges $E$ added by the solution.
        Gray circles are the neutral points $N$ added by the solution.
    }
    \label{fig:euclidean_decoding}
\end{figure}

\clearpage
\section{Additional Data}
\label{app:extra-data}

\begin{figure}[H]
    \centering
    \resizebox{\linewidth}{!}{
        \includegraphics{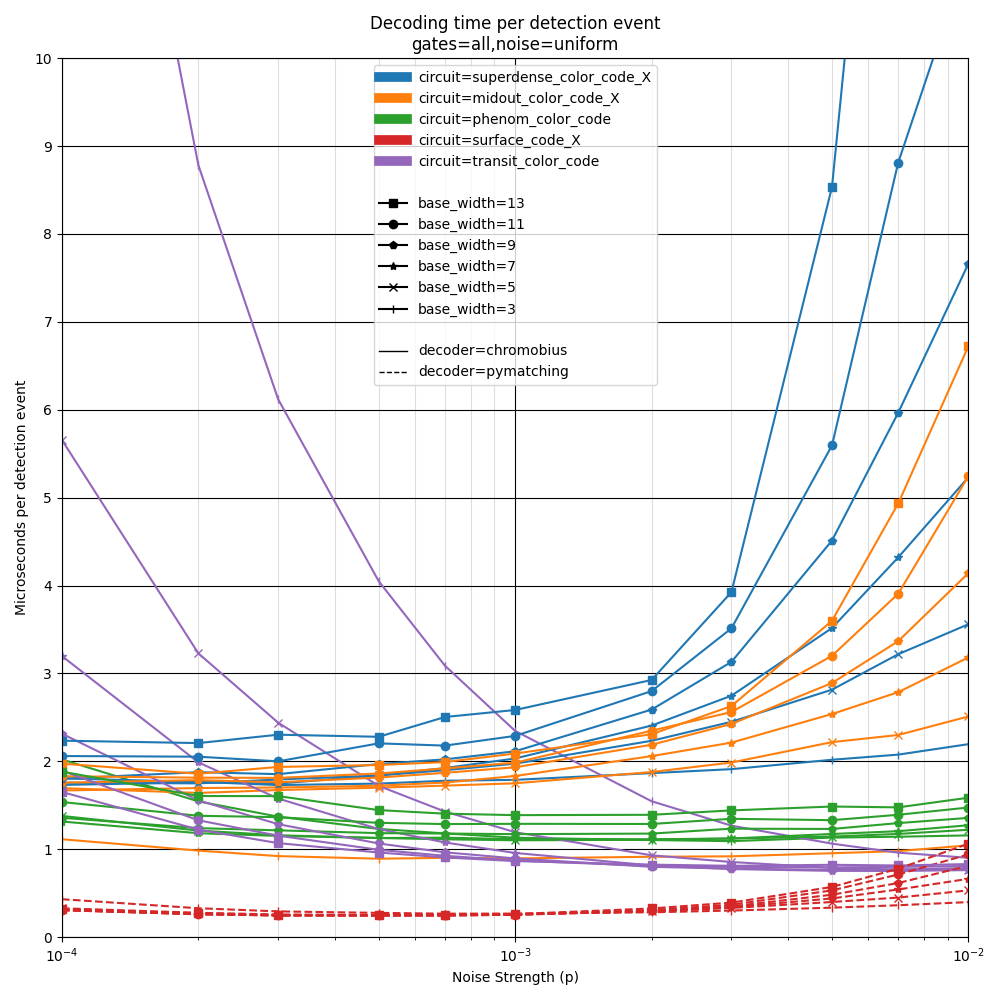}
    }
    \caption{
        Decoding time per detection event for all circuits, sizes, noise strengths, and decoders that were timed.
        The purple curves behave differently at low noise because, at low noise, decoding the code capacity color code is strongly dominated by fixed costs (rather than per-detection-event costs).
    }
    \label{fig:timing_d_full}
\end{figure}

\begin{figure}
    \centering
    \resizebox{\linewidth}{!}{
        \includegraphics{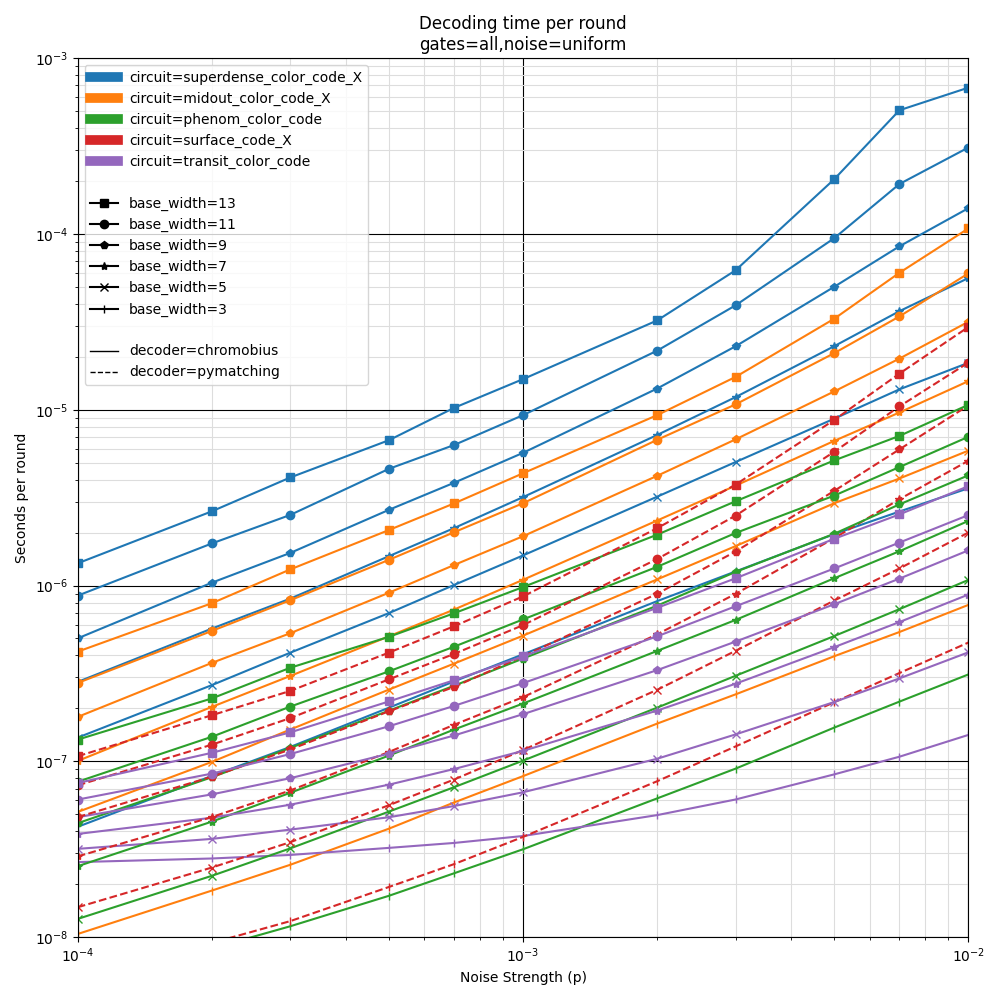}
    }
    \caption{
        Decoding time per round for all circuits, sizes, noise strengths, and decoders that were timed.
    }
    \label{fig:timing_r_full}
\end{figure}

\begin{figure}
    \centering
    \resizebox{0.48\linewidth}{!}{
        \includegraphics{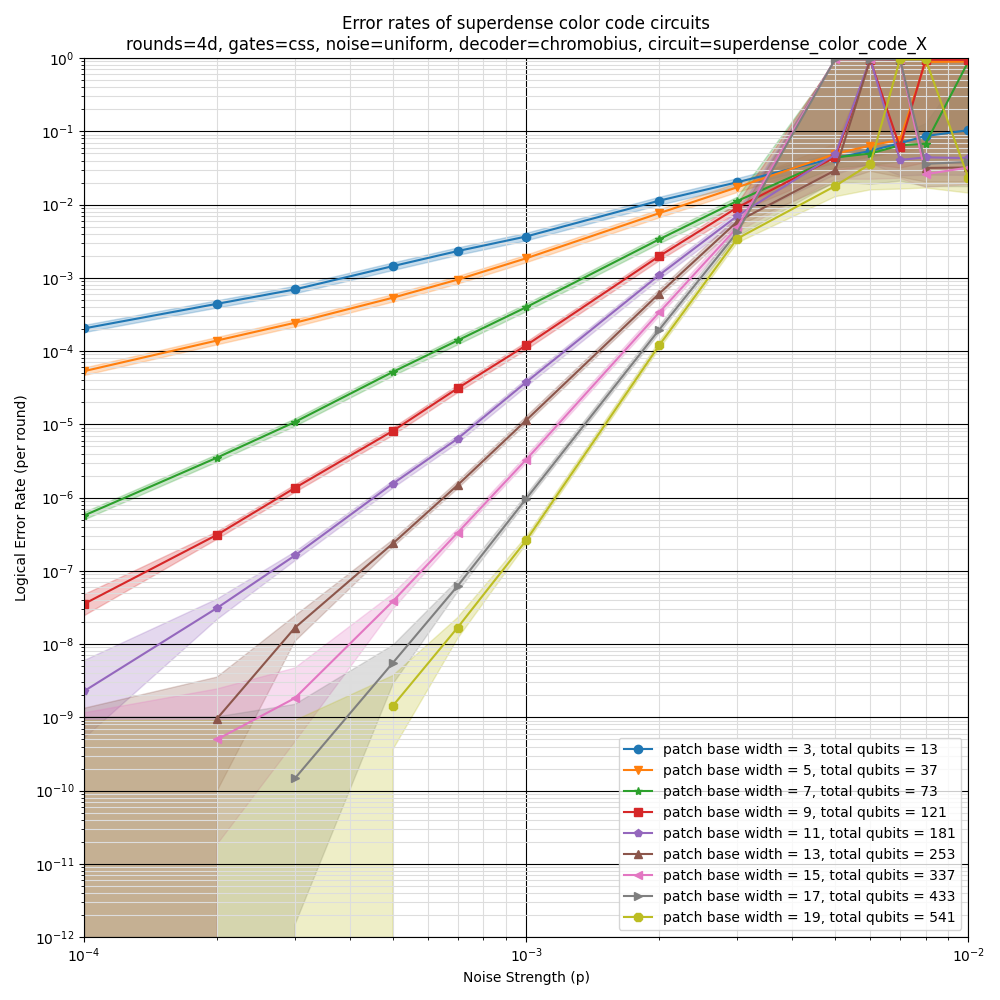}
    }
    \hfill
    \resizebox{0.48\linewidth}{!}{
        \includegraphics{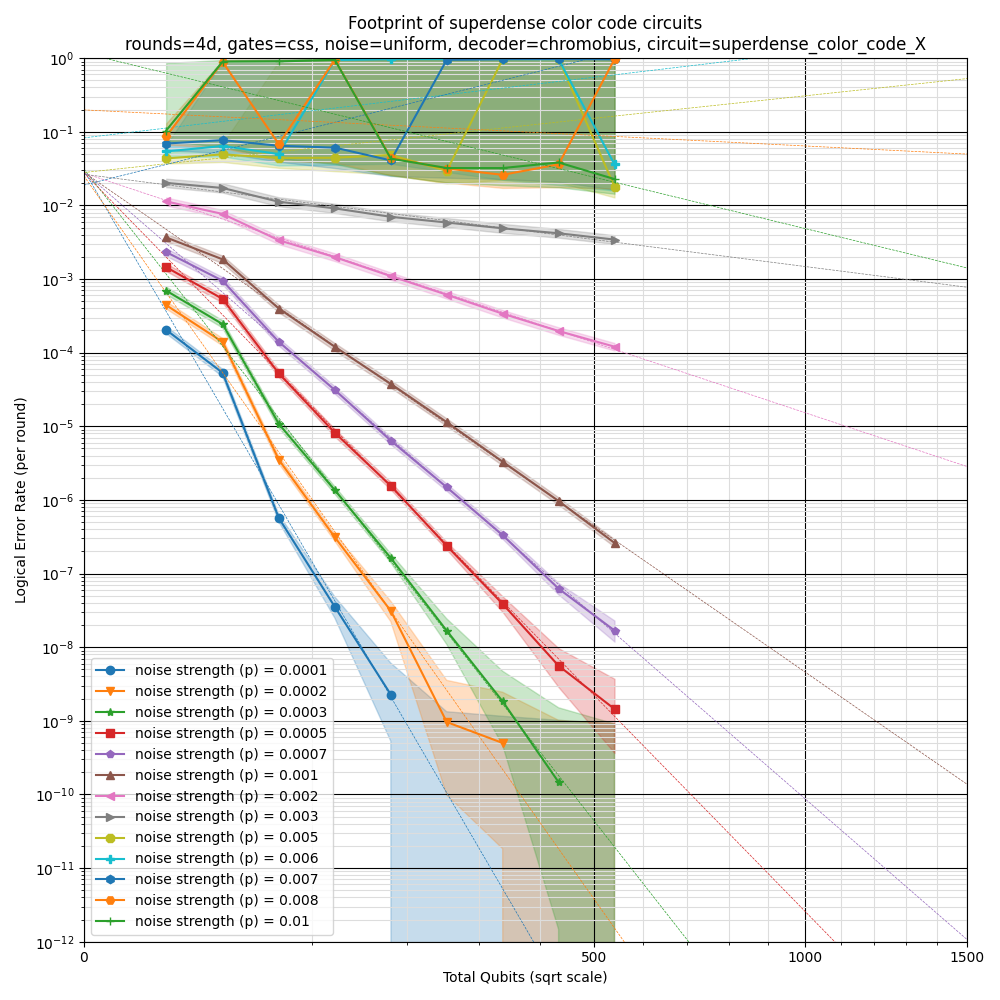}
    }
    \caption{
        Threshold plot (left) and footprint extrapolation plot (right) for X-basis superdense color code circuits.
        Highlights show hypotheses with likelihoods within a factor of 1000 of the max likelihood hypothesis, given the sampled data.
    }
    \label{fig:superdense-x}
\end{figure}

\begin{figure}
    \centering
    \resizebox{0.48\linewidth}{!}{
        \includegraphics{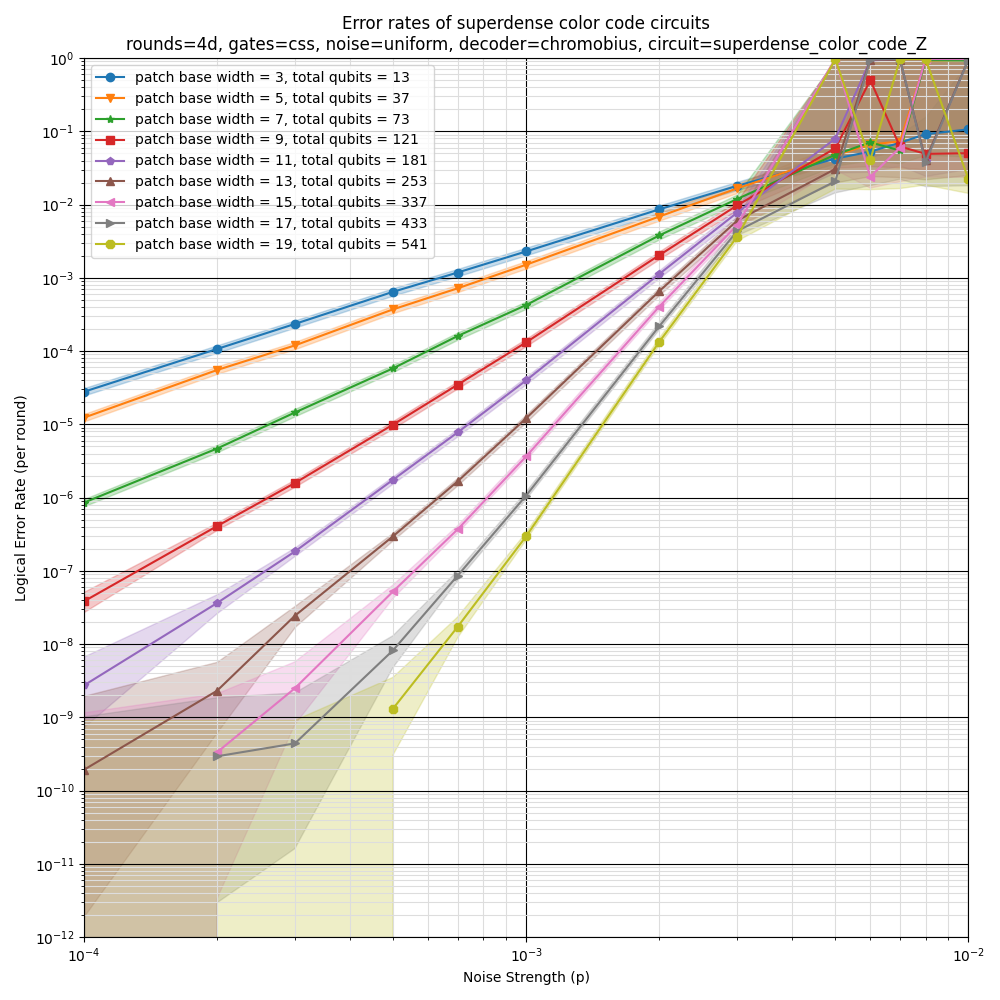}
    }
    \hfill
    \resizebox{0.48\linewidth}{!}{
        \includegraphics{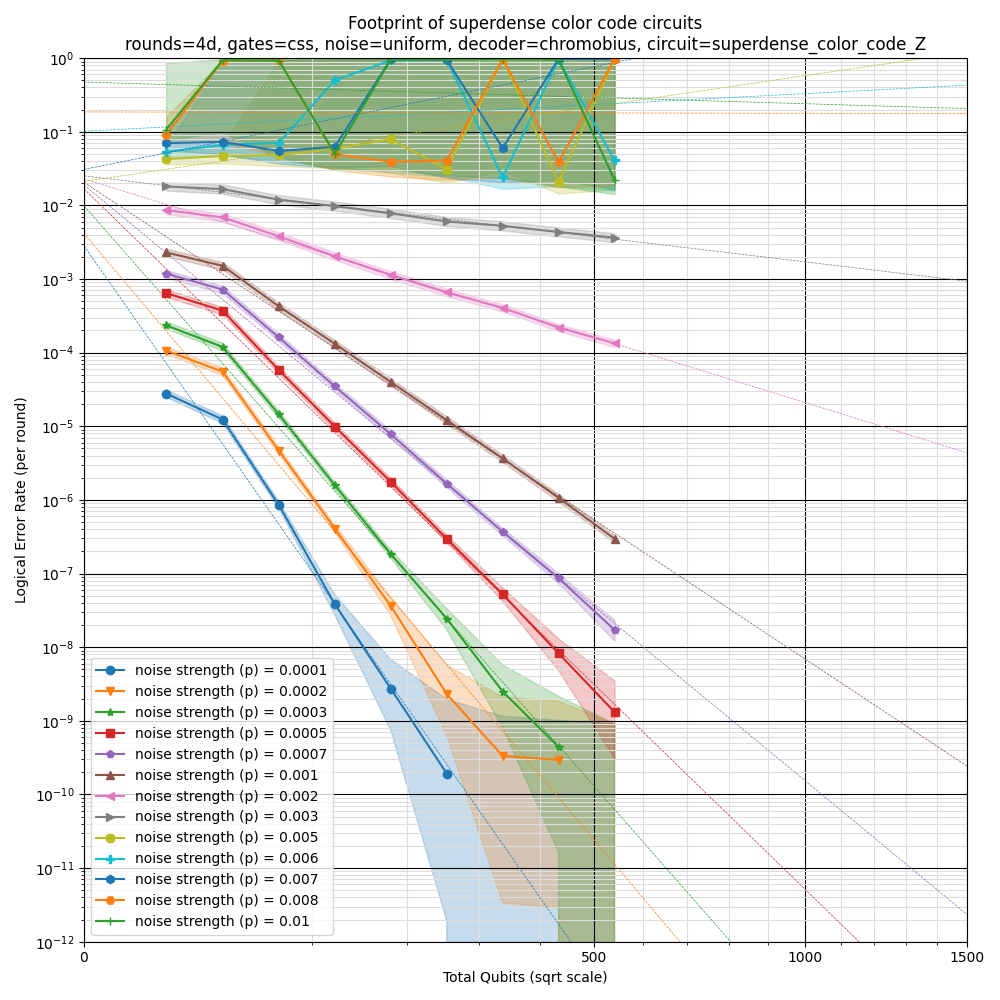}
    }
    \caption{
        Threshold plot (left) and footprint extrapolation plot (right) for Z-basis superdense color code circuits.
        Highlights show hypotheses with likelihoods within a factor of 1000 of the max likelihood hypothesis, given the sampled data.
    }
    \label{fig:superdense-z}
\end{figure}

\begin{figure}
    \centering
    \resizebox{0.48\linewidth}{!}{
        \includegraphics{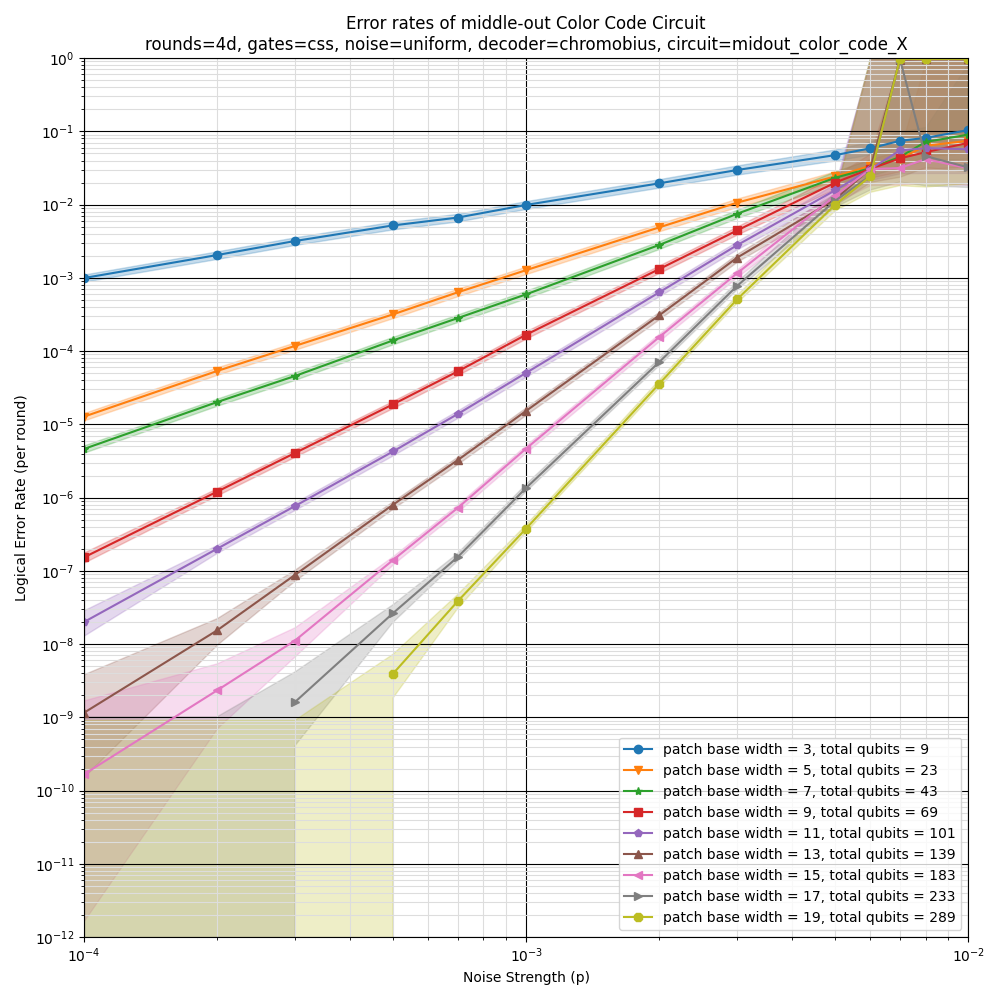}
    }
    \hfill
    \resizebox{0.48\linewidth}{!}{
        \includegraphics{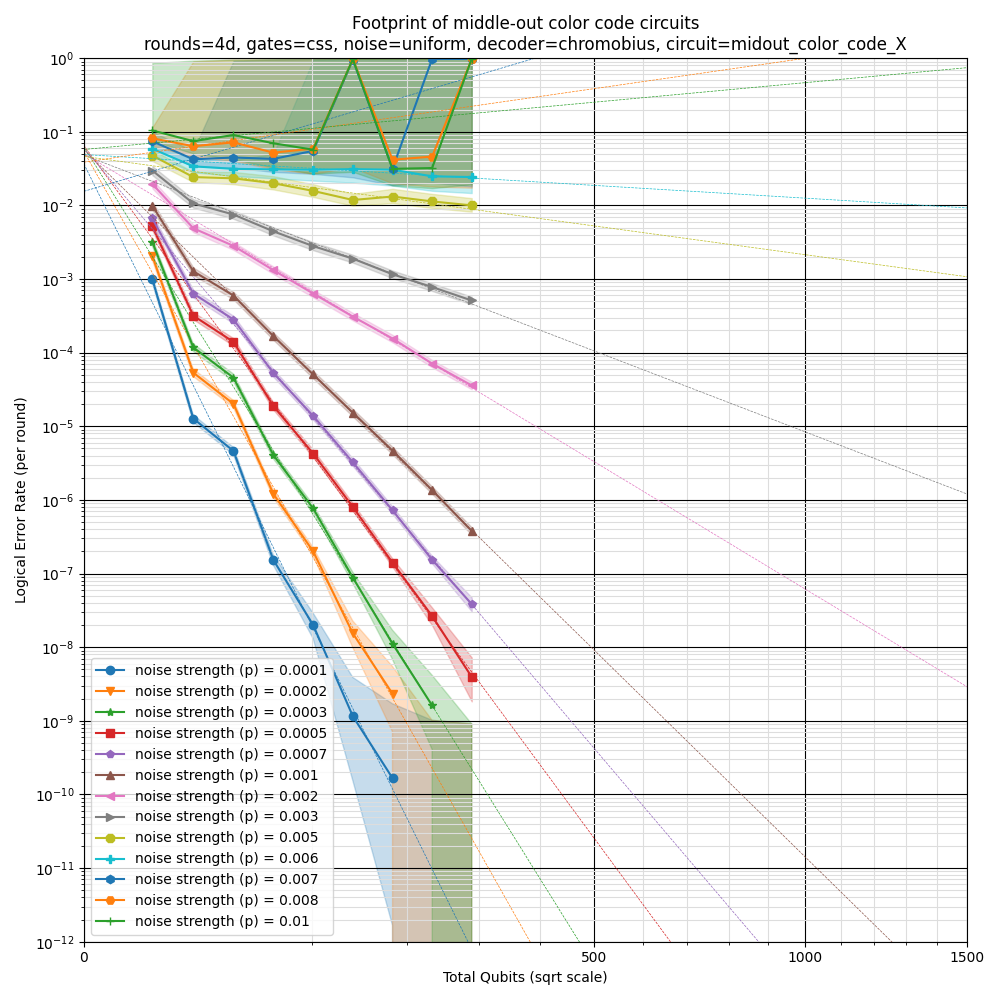}
    }
    \caption{
        Threshold plot (left) and footprint extrapolation plot (right) for X-basis middle-out color code circuits.
        Highlights show hypotheses with likelihoods within a factor of 1000 of the max likelihood hypothesis, given the sampled data.
    }
    \label{fig:midout-x}
\end{figure}

\begin{figure}
    \centering
    \resizebox{0.48\linewidth}{!}{
        \includegraphics{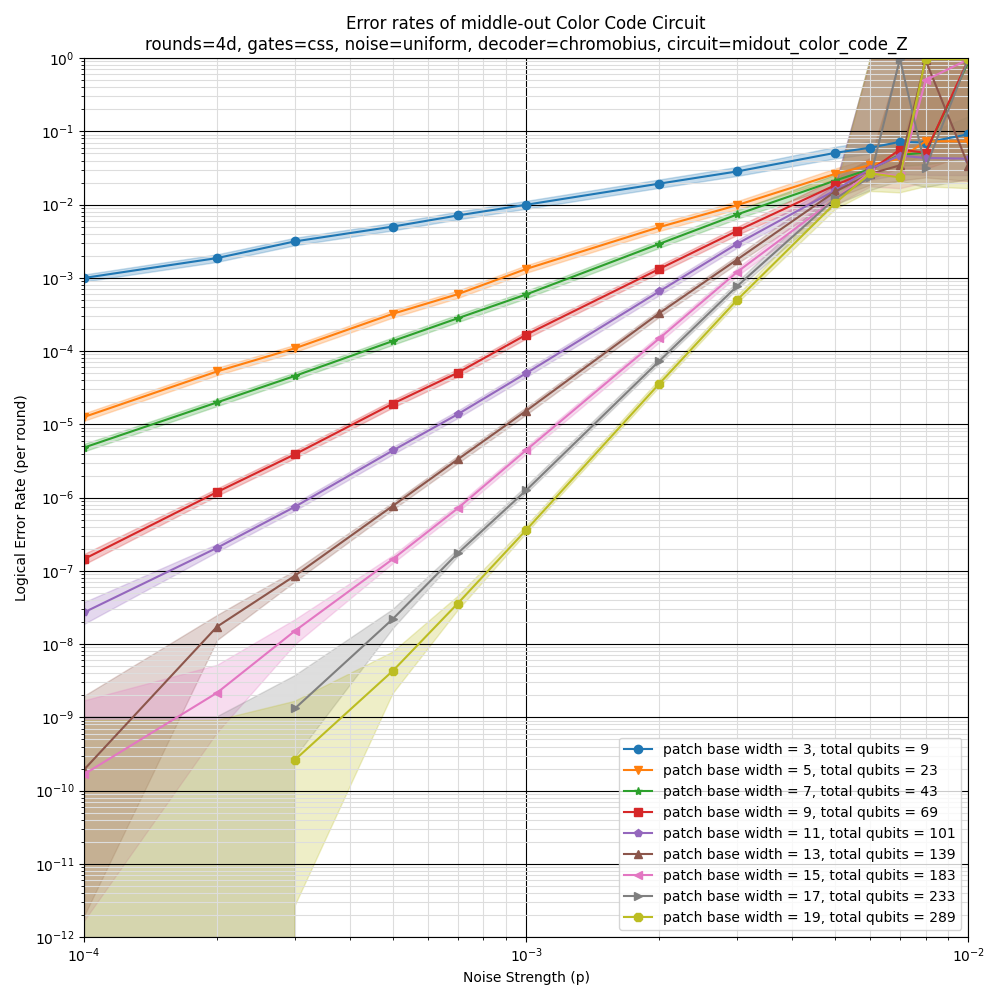}
    }
    \hfill
    \resizebox{0.48\linewidth}{!}{
        \includegraphics{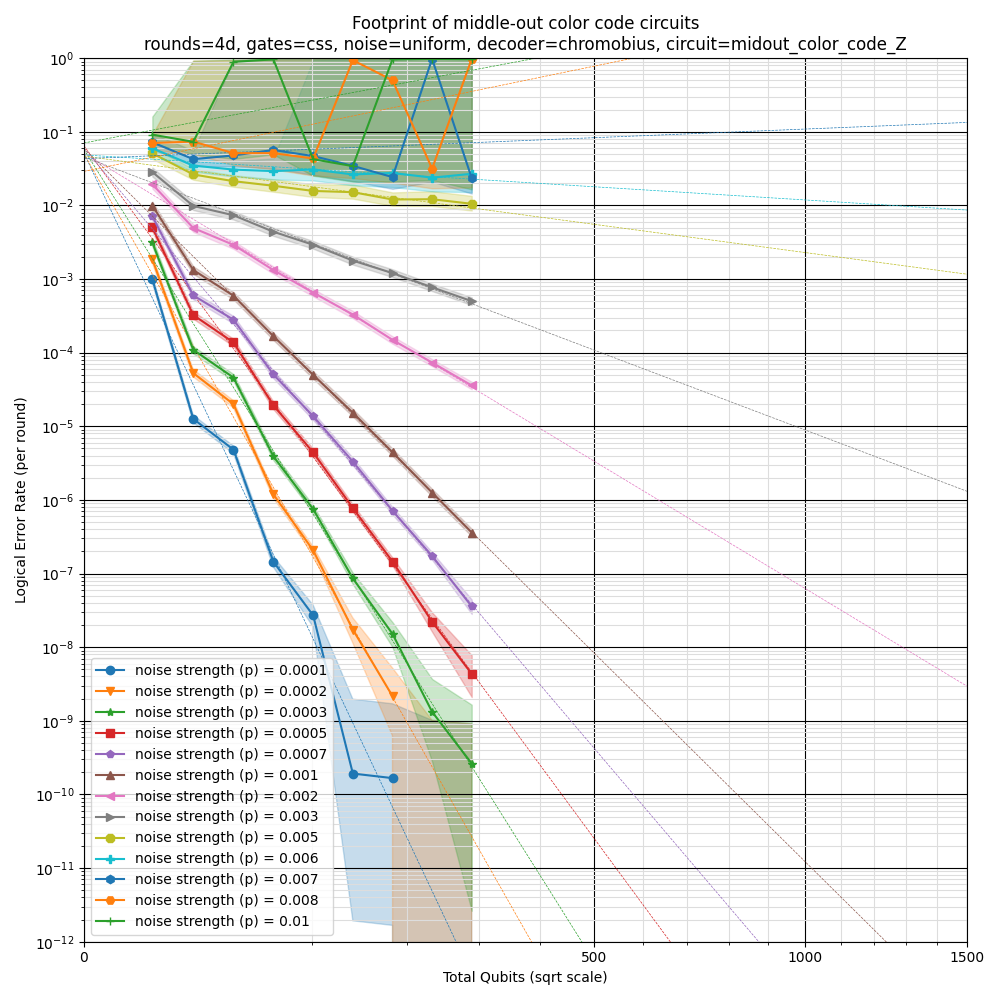}
    }
    \caption{
        Threshold plot (left) and footprint extrapolation plot (right) for Z-basis middle-out color code circuits.
        Highlights show hypotheses with likelihoods within a factor of 1000 of the max likelihood hypothesis, given the sampled data.
    }
    \label{fig:midout-z}
\end{figure}

\end{document}